\documentclass[a4paper,fleqn]{cas-sc}

\usepackage[numbers]{natbib}
\usepackage{hyperref}

\def\tsc#1{\csdef{#1}{\textsc{\lowercase{#1}}\xspace}}
\tsc{WGM}
\tsc{QE}
\tsc{EP}
\tsc{PMS}
\tsc{BEC}
\tsc{DE}

\begin{document}
\let\WriteBookmarks\relax
\def\textpagefraction{.001}

 \shorttitle{LAD vs GBR Methods for High-Speed Flows}

\shortauthors{Ritu Raj et~al.}

\title [mode = title]{Assessment of Gradient-based Reconstruction and Artificial Diffusivity Methods in Simulating High-Speed Compressible Flows} 

\tnotemark[1,2]


%
\author[1]{Ritu Raj Kumar}[type=editor,
                        auid=000,bioid=1]

\cormark[1]

\fnmark[1]

\ead{ae22d016@smail.iitm.ac.in}


\credit{Methodology, Conceptualisation, Investigation, Formal analysis, Original draft, Writing – review \& editing, Validation}

\affiliation[1]{organization={Indian Institute of Technology Madras},
    city={Chennai},
    postcode={600036}, 
    state={Tamil Nadu},
    country={India}}

\author[1]{Shivam Saini}
\credit{Conceptualisation, Methodology, Software}

\author[1]{Nagabhushana Rao Vadlamani}[%
   ]
\cormark[2]
\ead{nrv@iitm.ac.in}

\credit{Supervision, Resources, Funding, Project administration, original draft, Writing – review \& editing}


\author%
[2]
{Amareshwara Sainadh Chamarthi}

\affiliation[2]{organization={California Institute of Technology},
    city={Pasadena},
    country={USA}}

\credit{Methodology, Conceptualisation, Supervision, Writing – review}

\cortext[cor1]{Corresponding author}
\cortext[cor2]{Principal corresponding author}



\begin{abstract}
The two promising methods for capturing high-speed flows are local artificial diffusivity (LAD) and centralised gradient-based reconstruction (C-GBR), the former being computationally economical and the latter being more robust and stable but expensive. While the LAD approach captures discontinuities by adding artificial fluid transport coefficients, C-GBR employs a wave appropriate discontinuity sensor to obtain cleaner results and utilises the HLLC approximate Riemann solver for computing inviscid fluxes. The efficacy of these schemes is initially demonstrated in single-species 1D and 2D test cases. Moreover, the shock-capturing capability is assessed for 3D supersonic and hypersonic turbulent boundary layers. The accuracy of LAD predictions is comparable to that of C-GBR for the test case of a supersonic turbulent boundary layer. From the stability front, all the simulations are found to be stable with the C-GBR scheme, while the LAD-based simulations are observed to abruptly diverge for supersonic and hypersonic flows over compression corners with stronger shocks and larger flow separations. From the computational front, the LAD-based schemes are $1.17-2.32\times$ faster than the monotonicity-preserving explicit/implicit C-GBR schemes. A hybrid approach leveraging the strengths of LAD and GBR schemes is proposed as a promising solution for high-speed turbulent flows with strong shock–boundary layer interactions. The efficacy of the hybrid LAD-GBR solver is demonstrated for the compressible triple-point and supersonic compression ramp test cases. For the M2.9, $24^{\circ}$ case, the hybrid solver was stable and achieved a notable $1.67\times$ speed-up over the C-GBR scheme.

\end{abstract}


\begin{highlights}
 \item Evaluated performance of LAD and C-GBR schemes on high-speed flows.
 \item LAD achieves $1.17-2.32 \times$ speedup over C-GBR.
 \item Unlike LAD, C-GBR schemes are stable and capture strong discontinuities.
 \item Proposed a stable and economical hybrid framework that combines LAD and GBR.
 \item  For the M2.9, $24^{\circ}$ case, hybrid solver is $1.67\times$ faster than C-GBR.
\end{highlights}


\maketitle

\section{Introduction}

The interaction between shock waves and turbulence is widely encountered in supersonic and hypersonic applications. In high-speed turbulent flows, two primary discontinuities are (a) a shock across which density, pressure, and the normal velocity are discontinuous and (b) a contact (material) discontinuity across which density and temperature are discontinuous. Capturing these sharp discontinuities while equally resolving turbulence scales is a challenging task. Numerical methods for high-speed turbulent flows face a fundamental trade-off: accurately resolving all relevant turbulent scales requires minimal numerical dissipation, whereas capturing shock waves demands the use of shock-capturing schemes that inherently introduce artificial dissipation to ensure numerical stability.

It is well known that central difference schemes are preferred over the upwind schemes in smooth regions of the flow. However, these central schemes can result in spurious oscillations when applied to sharp gradients and discontinuities. In the context of Large Eddy Simulation (LES), several schemes have been proposed in the literature to meet such conflicting requirements. These include variants of Essentially Non-Oscillatory schemes (Weighted ENO -WENO/Targeted ENO - TENO), Monotonicity-Preserving (MP) approach, hybrid WENO-central difference, artificial diffusivity, adaptive characteristic-based filter, and shock fitting \cite{johnsen2010assessment}. Artificial diffusivity-based schemes are attractive for discontinuity capturing owing to their algorithmic simplicity and reduced computational overhead. \textcolor{black}{Cook} \cite{cook2007artificial} proposed a methodology to capture discontinuities in the flows involving shocks, turbulence, and mixing. The 1-D formulation of artificial diffusivity coefficients \cite{cook2007artificial} was further extended to multi-dimensional curvilinear coordinates by Kawai \& Lele \cite{kawai2008localized}. The method is based on high-wavenumber damping, which adds localised artificial fluid transport coefficients such as dynamic viscosity, bulk viscosity, thermal conductivity, and species diffusivity to the physical transport fluid coefficients across the discontinuity. In this method, the discontinuity is smeared over a numerically resolvable scale. Kawai et al. \cite{kawai2010assessment} have investigated the performance of different local artificial diffusivity (LAD) models with compact central difference schemes for large eddy simulation of compressible turbulent flows. The authors proposed a novel switching function (shock sensor), which combines a negative dilatation and Ducros-type sensor to model the artificial bulk viscosity. Further, \textcolor{black}{Olson and Lele} \cite{olson2013directional} developed a directional artificial fluid diffusivity for capturing shocks on the grids with a high aspect ratio (AR). Kumar and Vadlamani \cite{kumar2022scalar} showed that directional LAD is more efficient than scalar LAD on highly stretched grids with large aspect ratios. The key advantages of artificial diffusivity methods include its ease of implementation, low computational cost, localisation of artificial fluid transport coefficients, and lack of a weighting/hybrid scheme. An alternate shock-capturing strategy involves WENO-based schemes, which were first introduced by Liu et al. \cite{liu1994weighted} and later improved by Jiang et al. \cite{jiang1996efficient}. Nevertheless, for shock wave turbulent boundary layer interactions, the classical WENO schemes introduce excessive dissipation, damping small-scale turbulent structures. Subsequent developments have aimed to overcome these shortcomings by addressing aspects such as improving the formal order of accuracy at the critical points, optimising dispersion and reducing non-linear dissipation, and optimising spectral properties \cite{henrick2005mapped, borges2008improved, hu2010adaptive, hu2011scale, mullenix2011bandwidth, hu2012dispersion, arshed2013minimizing}. Recently, \textcolor{black}{ Fernández-Fidalgo et al.} \cite{fernandez2018posteriori} proposed a high-order hybrid method consisting of explicit central schemes for smooth regions and a compact WENO scheme across discontinuities. The authors have used \textit{a posteriori} criteria to detect points in the solution domain with negative density and pressure and then recalculate the numerical fluxes using WENO and positivity-preserving flux limiters. Moreover, several studies in the literature \cite{volpiani2018effects, volpiani2020effects, bernardini2021streams, adler2019flow} have used a modified form of the Ducros shock sensor \cite{ducros1999large} with a threshold cutoff to switch between the WENO and the central difference schemes. Despite the improvements, numerical instabilities can arise when the schemes are applied to high Mach number flows and near-vacuum conditions. This issue can be mitigated through the use of positivity-preserving limiters along with a strong stability-preserving (SSP) Runge–Kutta method for time marching  \cite{fu2023review}. Unlike the WENO method, TENO schemes maintain minimal numerical dissipation within the desired wave-number range while preserving strong shock-capturing ability. Recent studies \cite{fu2023review, fu2016family, fu2019very}  have proposed TENO schemes with optimised spectral properties and order of accuracy as high as 10$^{th}$ order for all speed regime gas dynamics and turbulence. However, stencil selection and weighting strategies can introduce significant computational overhead in large-scale simulations.

Recently, Chamarthi \cite{chamarthi2023gradient} proposed a high-accuracy gradient-based reconstruction approach for simulating high-speed flows. The proposed algorithm efficiently uses the computed gradient between the inviscid and viscous schemes. The methodology uses a monotonicity-preserving (MP) scheme in characteristic space to capture shock and material discontinuities. The fluxes are estimated using an approximate HLLC (Harten–Lax–van Leer contact) Riemann solver. The proposed nonlinear monotonicity preserving schemes (Explicit 6$^{th}$ order: MEG6, Implicit 4$^{th}$ order: MIG4) are shown to have superior spectral properties compared to the TENO5 schemes of Fu \cite{fu2019low} for a wide range of testcases. Chandravamsi et al. \cite{chandravamsi2023application} extended the gradient-based reconstruction approach of Chamarthi \cite{chamarthi2023gradient} for generalised curvilinear domains. Furthermore, Chamarthi \cite{chamarthi2023wave} proposed a wave-appropriate discontinuity sensor to selectively treat the various waves associated with compressible Euler equations. A combination of pressure and Ducros-based shock sensor is employed for the treatment of acoustic and shear/vortical characteristic waves, thus making the numerical dissipation more localised. An MP limiter is employed on the entropy/contact characteristic waves to mitigate oscillations. Hoffmann et al. \cite{hoffmann2024centralized} highlighted that the schemes of Chamarthi \cite{chamarthi2023wave} based on upwind linear interpolation (left- and right-state biased interpolations) can lead to additional unphysical dissipation even with selective limiting of the primitive variables in the characteristic space. The authors proposed improvements to these schemes using a centralised interpolation (averaging left- and right-state biased interpolations) for all the characteristic waves except the acoustic wave.

In summary, several methods have been proposed for shock capturing, each with notable limitations. The local artificial diffusivity (LAD) and hybrid WENO approaches are less dissipative, easy to implement, and low cost, but shock sensor localisation is not universal and may need case-specific tuning. The family of WENO schemes inherently suffers from numerical dissipation and instabilities, requiring bandwidth-optimised variants and positivity-preserving limiters. TENO schemes preserve strong shock-capturing with minimal dissipation, although stencil selection and weighting strategies can lead to increased computational overhead for large-scale simulations. The centralised gradient-based reconstruction (C-GBR) achieves spectral properties superior to those of TENO5 but incurs a higher cost than LAD methods. The primary objective of this paper is to assess the performance of local artificial diffusivity \cite{kawai2010assessment} and centralised gradient-based reconstruction methods \cite{hoffmann2024centralized} to capture compressible turbulent flows with shocks and contact discontinuities. In particular, we evaluate the performance of these numerical approaches when coupled with a high-order in-house solver, COMPSQUARE \cite{vadlamani2018distributed}, through several inviscid and viscous test cases. Additionally, we demonstrate that all the simulations are stable with C-GBR schemes, while LAD results in intermittent oscillations and diverges abruptly in the presence of stronger discontinuities. To this end, a hybrid LAD–GBR solver is proposed that combines the computational efficiency of LAD with the robustness of GBR for simulating high-speed flows. The efficacy of the hybrid solver is validated on a few test cases in which the LAD schemes failed to converge.

The remainder of this paper is organised as follows. In Section \ref{numerics}, the governing equations of the in-house solver COMPSQUARE are discussed. This is followed by a brief description of the LAD and C-GBR frameworks for simulating high-speed compressible flows. Section \ref{results} demonstrates the efficacy of the framework, highlighting accuracy, robustness, and performance across a variety of test cases. Section \ref{overview} compares the pseudocodes of both approaches and their impact on computational efficiency. Section \ref{hybrid} illustrates a hybrid solver approach that combines the LAD and GBR schemes and evaluates its efficacy. Conclusions are provided in Section \ref{summary}.




\section{Numerical Framework} 
\label{numerics}


\subsection{Governing equations}
The in-house solver COMPSQUARE \cite{vadlamani2018distributed} solves the unsteady three-dimensional compressible form of the Navier-Stokes equation in generalised curvilinear coordinates. These equations are expressed in conservative form as follows:

\begin{equation} \label{eq:3.1.1}
   \frac{\partial}{\partial t} \left( \frac{\mathbf{Q}}{J} \right)
 + \frac{\partial \mathbf{F}^{c}}{\partial \xi} + \frac{\partial \mathbf{G}^{c}}{\partial \eta} + \frac{\partial \mathbf{H}^{c}}{\partial \zeta} + \frac{\partial \mathbf{F}^{v}}{\partial \xi} + \frac{\partial \mathbf{G}^{v}}{\partial \eta} + \frac{\partial \mathbf{H}^{v}}{\partial \zeta}= 0   
\end{equation}

where $Q=[\rho,\rho u, \rho v, \rho w, E]^T$ is the vector of dependent conservative variables, ${\mathbf{F}^c}, {\mathbf{G}^c} ,$ and ${\mathbf{H}^c}$ are the convective flux vectors and $\mathbf{F}^{v}, \mathbf{G}^{v}$ and $\mathbf{H}^{v}$ are viscous flux vectors. The transformation Jacobian is represented as $J = \partial(\xi,\eta,\zeta)/\partial(x,y,z)$, where $\xi$, $\eta$, $\zeta$ are the computational coordinates and $t$ is the time. The convective or inviscid flux vectors are given as:

\begin{subequations}
\begin{equation}
\begin{gathered}
  \mathbf{F}^c = 
  \begin{pmatrix}
  \rho \hat{U}\\
  \rho u\hat{U} + \hat{\xi}_{x}p  \\
  \rho v\hat{U} + \hat{\xi}_{y}p  \\
  \rho w\hat{U} + \hat{\xi}_{z}p \\
  (\rho E + p)\hat{U}    
  \end{pmatrix}
  ,\quad \mathbf{G}^c = 
  \begin{pmatrix}
  \rho \hat{V}\\
  \rho u\hat{V} + \hat{\eta}_{x}p  \\
  \rho v\hat{V} + \hat{\eta}_{y}p  \\
  \rho w\hat{V} + \hat{\eta}_{z}p \\
  (\rho E + p)\hat{V}  
  \end{pmatrix}
  ,\quad \mathbf{H}^c = 
  \begin{pmatrix}
  \rho \hat{W}\\
  \rho u\hat{W} + \hat{\zeta}_{x}p  \\
  \rho v\hat{W} + \hat{\zeta}_{y}p  \\
  \rho w\hat{W} + \hat{\zeta}_{z}p \\
  (\rho E + p)\hat{W} 
  \end{pmatrix}
\end{gathered}
\end{equation}

\begin{equation}
\hat{U} = \hat{\xi}_{x}u + \hat{\xi}_{y}v + \hat{\xi}_{z}w, \quad
\hat{V} = \hat{\eta}_{x}u + \hat{\eta}_{y}v + \hat{\eta}_{z}w, \quad
\hat{W} = \hat{\zeta}_{x}u + \hat{\zeta}_{y}v + \hat{\zeta}_{z}w
\end{equation}
\end{subequations}

 In the preceding expression, $\rho$ is the density, $u$, $v$, $w$ as cartesian velocity components, $p$ is the pressure, $T$ is the temperature, $E = e + (u^2+v^2+w^2)/2$ is the total energy per unit volume, $e=p/[\rho(\gamma-1)]$ is the specific internal energy, and $\gamma$ is the specific heat ratio. The primitive variables are normalized with reference quantities such as $\rho_{\infty}$, $U_{\infty}$, $T_{\infty}$ and pressure is normalized with $\rho_{\infty}u_{\infty}^{2}$. The metric terms are normalised with the Jacobian ($\hat{\xi}_x = \xi_x/J$). The contravariant velocities are $\hat{U}, \hat{V}, \ \text{and}\ \hat{W}$. The viscous flux vectors $\mathbf{F}^{v}, \mathbf{G}^{v}$ and $\mathbf{H}^{v}$ are defined as:

\begin{subequations}
\begin{equation}
\begin{gathered}
  \hat{\mathbf{F}}_v = -
  \begin{bmatrix}
    0 \\
    \hat{\xi}_{x}\tau_{xx} + \hat{\xi}_{y}\tau_{xy} + \hat{\xi}_{z}\tau_{xz}  \\
    \hat{\xi}_{x}\tau_{yx} + \hat{\xi}_{y}\tau_{yy} + \hat{\xi}_{z}\tau_{yz}  \\
    \hat{\xi}_{x}\tau_{zx} + \hat{\xi}_{y}\tau_{zy} + \hat{\xi}_{z}\tau_{zz}  \\
    \hat{\xi}_{x}\varphi_{x} + \hat{\xi}_{y}\varphi_{y} + \hat{\xi}_{z}\varphi_{z}  
  \end{bmatrix}
  ,\quad \hat{\mathbf{G}}_v = -
  \begin{bmatrix}
    0 \\
    \hat{\eta}_{x}\tau_{xx} + \hat{\eta}_{y}\tau_{xy} + \hat{\eta}_{z}\tau_{xz}  \\
    \hat{\eta}_{x}\tau_{yx} + \hat{\eta}_{y}\tau_{yy} + \hat{\eta}_{z}\tau_{yz}  \\
    \hat{\eta}_{x}\tau_{zx} + \hat{\eta}_{y}\tau_{zy} + \hat{\eta}_{z}\tau_{zz}  \\
    \hat{\eta}_{x}\varphi_{x} + \hat{\eta}_{y}\varphi_{y} + \hat{\eta}_{z}\varphi_{z}  
  \end{bmatrix}
  ,\quad \hat{\mathbf{H}}_v = -
  \begin{bmatrix}
    0 \\
    \hat{\zeta}_{x}\tau_{xx} + \hat{\zeta}_{y}\tau_{xy} + \hat{\zeta}_{z}\tau_{xz}  \\
    \hat{\zeta}_{x}\tau_{yx} + \hat{\zeta}_{y}\tau_{yy} + \hat{\zeta}_{z}\tau_{yz}  \\
    \hat{\zeta}_{x}\tau_{zx} + \hat{\zeta}_{y}\tau_{zy} + \hat{\zeta}_{z}\tau_{zz}  \\
    \hat{\zeta}_{x}\varphi_{x} + \hat{\zeta}_{y}\varphi_{y} + \hat{\zeta}_{z}\varphi_{z}  
  \end{bmatrix}
\end{gathered}
\end{equation}

\begin{equation}
\label{tauij}
\tau_{ij} = \hat{\mu} \left( \frac{\partial \xi_k}{\partial x_j} \frac{\partial u_i}{\partial \xi_k} 
+ \frac{\partial \xi_k}{\partial x_i} \frac{\partial u_j}{\partial \xi_k}\right)
+ \left(\beta - \frac{2}{3} \hat{\mu}\right)  \frac{\partial \xi_l}{\partial x_k} \frac{\partial u_k}{\partial \xi_l}\delta_{ij} , \quad
\varphi_i = u_j \tau_{ij} + \hat{\kappa} \frac{\partial T}{\partial x_i}
\end{equation}
\end{subequations}

where $\tau_{ij}$ is the viscous stress tensor and $\varphi_i$ is defined as the thermal stress term. The normalised dynamic viscosity is $\hat{\mu}=\mu/\text{Re}$, where $\text{Re}=\rho_{\infty}u_{\infty}L_{\text{ref}}/\mu_{\infty}$ and $\beta$ is the bulk viscosity coefficient. Following Stokes' hypothesis, the bulk viscosity coefficient ($\beta = \lambda + 2\hat{\mu}/3$) is assumed as negligible. $\hat{\kappa} = \hat{\mu}/[(\gamma-1)\text{Pr}{\text{M}}^2_\infty]$ is the normalised thermal conductivity, where $Pr$ is the Prandtl number. Sutherland law is used for calculating dynamic viscosity, and the equation of state $p=\rho T/(\gamma \text{M}^2_\infty)$ is invoked to close the governing equations. The in-house solver utilises high-order central difference schemes (explicit/compact) with a low-pass implicit filter \cite{gaitonde1998high} to suppress the dispersion associated with high frequencies. The governing equations are marched in time with an explicit fourth-order Runge-Kutta scheme (RK4,4) or a third-order explicit total variation diminishing (TVD) Runge-Kutta scheme.

\subsection{Local Artificial Diffusivity (LAD) Method} 

The artificial diffusivity scheme used in this study is based on a numerical framework proposed by Kawai et al. \cite{kawai2010assessment}. Shock and material discontinuities are captured by adding grid-dependent artificial fluid transport coefficients ($\mu^*, \beta^*, \kappa^*$) to the physical fluid transport coefficients as

\begin{equation}
   \mu =\mu_f+\mu^*, \quad \beta =\beta_f+\beta^*, \quad \kappa =\kappa_f+\kappa^* 
\end{equation}

 where, $\mu$ is the dynamic viscosity, $\beta$ is the bulk viscosity, and $\kappa$ is the thermal conductivity. An attractive feature of this method is the automatic deactivation of the artificial transport coefficients in smooth, well-resolved flow regions while providing dissipation in the non-smooth regions of the flow. The formulation of artificial diffusivity coefficients is given as,

\begin{equation}
   \mu^*=C_\mu \overline{\rho\left|\sum_{l=1}^3 \frac{\partial^4 S}{\partial \xi_l^4} \Delta \xi_l^4 \Delta_{l, \mu}^2\right|}, \quad  \kappa^*=C_\kappa \overline{\frac{\rho c_s}{T}\left|\sum_{l=1}^3 \frac{\partial^4 e}{\partial \xi_l^4} \Delta \xi_l^4 \Delta_{l, \kappa}\right|}
\end{equation}

\begin{equation}
    \beta^*=C_\beta \overline{\rho f_{s w}\left|\sum_{l=1}^3 \frac{\partial^4 \nabla\cdot\textbf{u}}{\partial \xi_l^4} \Delta \xi_l^4 \Delta_{l, \beta}^2\right|}, \quad f_{sw} = \frac{H(-\nabla\cdot\textbf{u})(\nabla\cdot\textbf{u})^2}{(\nabla \cdot \textbf{u})^{2} + (\nabla\times\textbf{u})^{2} + \epsilon}, 
\end{equation}

where, $S$ is the strain rate tensor, and $c_s$ is the sound speed. The dimensionless user-specified coefficients are $C_\mu = 0.002 $, $C_\beta$ = 1.75, and $C_\kappa$ = 0.01. The generalized coordinates denoted by $\xi_{l}$, where $l$ = 1, 2, and 3 represent the $\xi$, $\eta$ and $\zeta$ directions, respectively. $H(\cdot)$ is a Heaviside function, and $\epsilon$ is a small positive constant of magnitude $10^{-32}$ to prevent division by zero. $f_{sw}$ is the switching function, a combination of negative dilatation $H(-\nabla\cdot u)$ and Ducros-type sensor \cite{ducros1999large} that localises the artificial bulk viscosity across the shock waves. The over-bar denotes an approximate truncated-Gaussian filter (see \cite{kawai2008localized} for further details on the filter operation). The $\Delta\xi_{l}$ and $\Delta_{l} \mathbf{\cdot}$ are the grid spacings in the computational and physical space, where $\Delta\xi_{l}$ = 1. The grid spacing $\Delta_{l} \mathbf{\cdot}$ for each of the artificial diffusivity coefficients is defined as  

\begin{equation*}
\Delta _{l_,\mu} = \left| \Delta \mathbf{x}_l \right|, \quad
\Delta _{l_,\beta} = \left| \Delta \mathbf{x}_l \cdot \frac{\boldsymbol{\nabla} \mathbf{\rho}}{|\boldsymbol{\nabla} \mathbf{\rho}|} \right|, \quad
\Delta _{l_,\kappa} = \left| \Delta \mathbf{x}_l \cdot \frac{\boldsymbol{\nabla} \mathbf{e}}{|\boldsymbol{\nabla} \mathbf{e}|} \right|
\end{equation*}

where, $\mathbf{\Delta x}_{l}  = (0.5(x_{i+1}-x_{i-1}), 0.5(y_{i+1}-y_{i-1}), 0.5({z_{i+1}-z_{i-1}}))^{T}$, is the local displacement vector along the $\xi_{l}$ direction, with the nodes in the $\xi_{l}$ direction indexed by $i$.

\subsection{Centralised Gradient-Based Reconstruction (C-GBR) Method}

The high-accuracy centralised gradient-based reconstruction schemes used in this study are based on a numerical approach of  Chandravamsi et al. \cite{chandravamsi2023application} and Hoffman et al. \cite{hoffmann2024centralized}. The conservative variables are stored at node locations, while fluxes are computed at cell interfaces (half-locations) using finite difference approximations. The flux derivatives at the node are evaluated using fluxes at half-locations, resulting in a conservative finite difference approach. A monotonicity-preserving sixth-order explicit scheme (MEG6) and fourth-order implicit scheme (MIG4) are used for spatial discretisation, and a third-order explicit total variation diminishing (TVD) Runge-Kutta is employed for time marching. For the convective flux, a characteristic transformation of primitive variables is exploited to get cleaner results without spurious oscillations in the regions of discontinuities. In particular, a wave appropriate discontinuity sensor is employed to selectively treat each characteristic variable [acoustic (1,5), entropy/contact (2) and shear/vortical (3,4)] for superior results. A centralised interpolation (averaging left- and right-state biased interpolations) is used for all the characteristic waves, \textbf{C}, except the acoustic wave and is expressed as

\begin{equation}
\mathbf{C}^{L}_{i+\frac{1}{2}} = \mathbf{C}^{R}_{i+\frac{1}{2}} = \mathbf{C}^{C}_{i+\frac{1}{2}} = \frac{1}{2} \left( \mathbf{C}^{L, \text{Linear}}_{i+\frac{1}{2}} + \mathbf{C}^{R, \text{Linear}}_{i+\frac{1}{2}} \right)
\end{equation}

where $L, R$ denote the left- and right-biased states and $(\cdot)^C$  denotes the centralised interpolation. The interpolation of left-biased characteristic variables is performed as

\begin{equation}
\mathbf{C}_{i+\frac{1}{2}, b}^L= \begin{cases}\text { if } b=1,5: & \begin{cases}\mathbf{C}_{i+\frac{1}{2}, b}^{L, \text { Non-Linear }} & \text { if } \overline{\Omega_{i, j, k}^d}>0.01, \\[10pt] \mathbf{C}_{i+\frac{1}{2}, b}^{L, \text { Linear }} & \text { otherwise },\end{cases} \\ \\

\text { if } b=2: & \begin{cases}\mathbf{C}_{i+\frac{1}{2}, b}^{L, \text { Non-Linear }} & \text { if }\left(\mathbf{C}_{i+\frac{1}{2}}^{L, \text { Linear }}-\mathbf{C}_i\right)\left(\mathbf{C}_{i+\frac{1}{2}}^{L, \text { Linear }}-\mathbf{C}_{i+\frac{1}{2}}^{L, M P}\right) \leq 10^{-20}, \\[10pt] \mathbf{C}_{i+\frac{1}{2}, b}^{C, \text { Linear }} & \text { otherwise },\end{cases} \\ \\

\text { if } b=3,4: & \begin{cases}\mathbf{C}_{i+\frac{1}{2}, b}^{L, \text { Non-Linear }} & \text { if } \overline{\Omega_{i, j, k}^d}>0.01, \\[10pt] \mathbf{C}_{i+\frac{1}{2}, b}^{C, \text { Linear }} & \text { otherwise. }\end{cases} \end{cases}
\end{equation}

The same interpolation is performed for the right-biased characteristic variables. The nonlinear terms, $(\cdot)^{L, \ \text{Non-Linear}}$, and the acoustic waves are computed using an upwind interpolation based on a fifth-order monotonicity-preserving (MP) scheme. Linear terms, $(\cdot)^{L, \ \text{Linear}}$, are computed using the MEG6 or MIG4 schemes. Entropy or contact (2) characteristic variables use an MP limiting criterion, while the acoustic (1) and shear or vortical (3,4) waves use a combination of pressure and Ducros-based threshold ($\Omega_{i, j, k}^d$) to effectively suppress oscillations induced by discontinuities. The present approach does not use the high-frequency alpha damping schemes for viscous flux discretisation as proposed by Chamarthi \cite{chamarthi2023gradient}. In addition, only MP-based limiting criteria are used for inviscid flow simulations. The convective numerical fluxes are computed using the HLLC approximate Riemann solver as follows

\begin{subequations}
\begin{equation}
\mathbf{F}^{HLLC}_{i \pm \frac{1}{2}} =
\begin{cases}
\mathbf{F}^L, & \text{if } 0 \leq S^L, \\
\mathbf{F}_*^L = \mathbf{F}_*^L + S^L(\mathbf{Q}_*^L - \mathbf{Q}^L) & \text{if } S^L \leq 0 \leq S_*, \\
\mathbf{F}_*^R = \mathbf{F}_*^R + S^R(\mathbf{Q}_*^R - \mathbf{Q}^R)& \text{if } S_* \leq 0 \leq S^R, \\
\mathbf{F}^R, & \text{if } 0 \geq S^R.
\end{cases}
\end{equation}

\begin{equation}
\begin{aligned}
S^L &= \min \left( U^L - c^L \sqrt{\widetilde{\xi}_x^2 + \widetilde{\xi}_y^2 + \widetilde{\xi}_z^2},\ \breve{U} - \check{c} \right), \quad
S^R = \max \left( U^R + c^R \sqrt{\widetilde{\xi}_x^2 + \widetilde{\xi}_y^2 + \widetilde{\xi}_z^2},\ \breve{U} + \check{c} \right), \\
S_* &= \frac{
    \rho^R U^R (S^R - U^R) - \rho^L U^L (S^L - U^L) + (p^L - p^R)\left( \widetilde{\xi}_x^2 + \widetilde{\xi}_y^2 + \widetilde{\xi}_z^2 \right)
}{
    \rho^R (S^R - U^R) - \rho^L (S^L - U^L)
}.
\end{aligned}
\end{equation}
\end{subequations}

where the terms $(\cdot)^{L,R}$ represent the left or right states. $S^L$, $S^R$, and $S_*$ are the left, right, and star wave speeds, respectively. The sound speed is denoted as $c$, the Roe-averaged variables are denoted with the $\breve{(\cdot)}$ symbol, and $\tilde{(\cdot)}$ indicates the cell interface grid metrics obtained using interpolation from node location. $U$ is the contravariant velocity, and \textbf{U}$^{L,R}$ is the conservative variable vector. The star state, \textbf{Q}$_*^K$, where $K=L,R$, is computed as follows.

\begin{equation}
\mathbf{Q}_*^K = \rho^K \left( \frac{S^K - U^K}{S^K - S_*} \right)
\left\{
\begin{array}{c}
1 \\
\displaystyle\frac{\widetilde{\xi}_x S_* + (\widetilde{\xi}_y^2 + \widetilde{\xi}_z^2) u^K - \widetilde{\xi}_y \widetilde{\xi}_x v^K - \widetilde{\xi}_z \widetilde{\xi}_x w^K}{\widetilde{\xi}_x^2 + \widetilde{\xi}_y^2 + \widetilde{\xi}_z^2} \\
\displaystyle\frac{\widetilde{\xi}_y S_* - \widetilde{\xi}_x \widetilde{\xi}_y u^K + (\widetilde{\xi}_x^2 + \widetilde{\xi}_z^2) v^K - \widetilde{\xi}_z \widetilde{\xi}_y w^K}{\widetilde{\xi}_x^2 + \widetilde{\xi}_y^2 + \widetilde{\xi}_z^2} \\
\displaystyle\frac{\widetilde{\xi}_z S_* - \widetilde{\xi}_x \widetilde{\xi}_z u^K - \widetilde{\xi}_y \widetilde{\xi}_z v^K + (\widetilde{\xi}_x^2 + \widetilde{\xi}_y^2) w^K}{\widetilde{\xi}_x^2 + \widetilde{\xi}_y^2 + \widetilde{\xi}_z^2} \\
\displaystyle E^K + (S_* - U^K) \left[ \frac{S_*}{\widetilde{\xi}_x^2 + \widetilde{\xi}_y^2 + \widetilde{\xi}_z^2} + \frac{p^K}{\rho^K (S^K - U^K)} \right]
\end{array}
\right\}
\end{equation}


\section{Results \& Discussions} \label{results}
The in-house solver COMPSQUARE is enhanced with local artificial diffusivity (LAD) and centralised gradient-based reconstruction (C-GBR) schemes. The implementation has been ported to the GPUs using OpenACC directives to accelerate computations. With the LAD approach, a sixth-order compact difference central scheme (LAD-C6) is used, while MIG4/MEG6 spatial discretisation schemes are used with the C-GBR approach. It is important to note that only $C_{\beta}$ and  $C_{\kappa}$ are employed for all the test cases with the LAD scheme. Moreover, the contribution of ($C_{\kappa}$) to the flow field has been found to have minimal effect on the solution. A Courant–Friedrichs–Lewy (CFL) number of 0.2 is used for all test cases unless stated otherwise. Subsequently, the efficacy of these methods has been verified using a diverse set of benchmark test cases as listed below
\begin{enumerate}
    \item One-dimensional test cases: Sod, Lax, Le Blanc and Shu-Osher problems
    \item Two-dimensional cases:
    \begin{enumerate}
    \item Inviscid test cases: Shock entropy wave 
    test, Richtmyer–Meshkov Instability, shock vortex interaction, Kelvin Helmholtz instability, Riemann problem, compressible triple point,  shock-bubble interaction
    \item Viscous test cases: Double periodic shear layer, 2D hypersonic compression ramp, 2D oblique shock boundary layer interaction
    \end{enumerate}
    \item Three-dimensional test cases: LES of supersonic turbulent boundary layer, shock wave turbulent boundary layer interaction at M2.9 ($24^{\circ}$) and M7.2 ($8^{\circ}$) compression ramp.
\end{enumerate}


\subsection{One-Dimensional Inviscid Test Cases}

\subsubsection*{Shock Tube Problems}

Two classical shock tube problems are considered: the Sod problem \cite{sod1978survey} and the Lax problem \cite{lax1954weak}, both widely used in the validation of compressible flow solvers. Both cases are simulated on a computational domain of $x=[0,1]$ using N = 200 uniformly spaced grid points. The initial conditions of the Sod shock problem and the Lax problem are given as  

\begin{equation}
(\rho, u, p)_{\text{Sod}} =
\begin{cases}
(1, 0, 1), & \text{if } 0 \leq x < 0.5, \\
(0.125, 0, 0.1), & \text{if } 0.5 \leq x \leq 1,
\end{cases}
\end{equation}

\begin{equation}
(\rho, u, p)_{\text{Lax}} =
\begin{cases}
(0.445, 0.698, 3.528), & \text{if } 0 \leq x < 0.5, \\
(0.5, 0, 0.571), & \text{if } 0.5 \leq x \leq 1,
\end{cases}
\end{equation}

The flow is simulated until $t=0.2$ units for the Sod problem and $t=0.14$ units for the Lax problem. The inflow boundary condition is set to the initial conditions, while variables are extrapolated at the outflow boundary.

\begin{figure}
    \centering
    \includegraphics{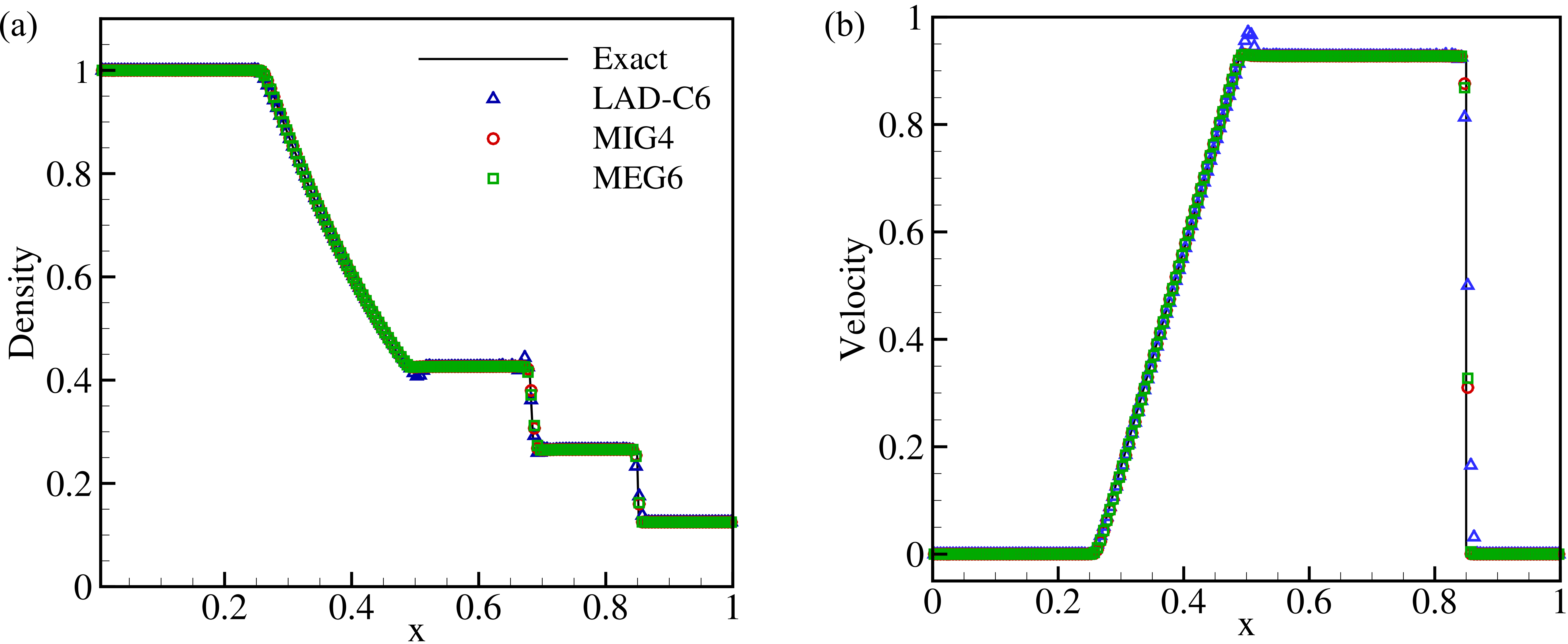}
    \caption{Plots of (a) density and (b) velocity profiles for the Sod shock tube problem compared against the exact solution.}
    \label{sodss}
\end{figure}

\begin{figure}
    \centering
    \includegraphics{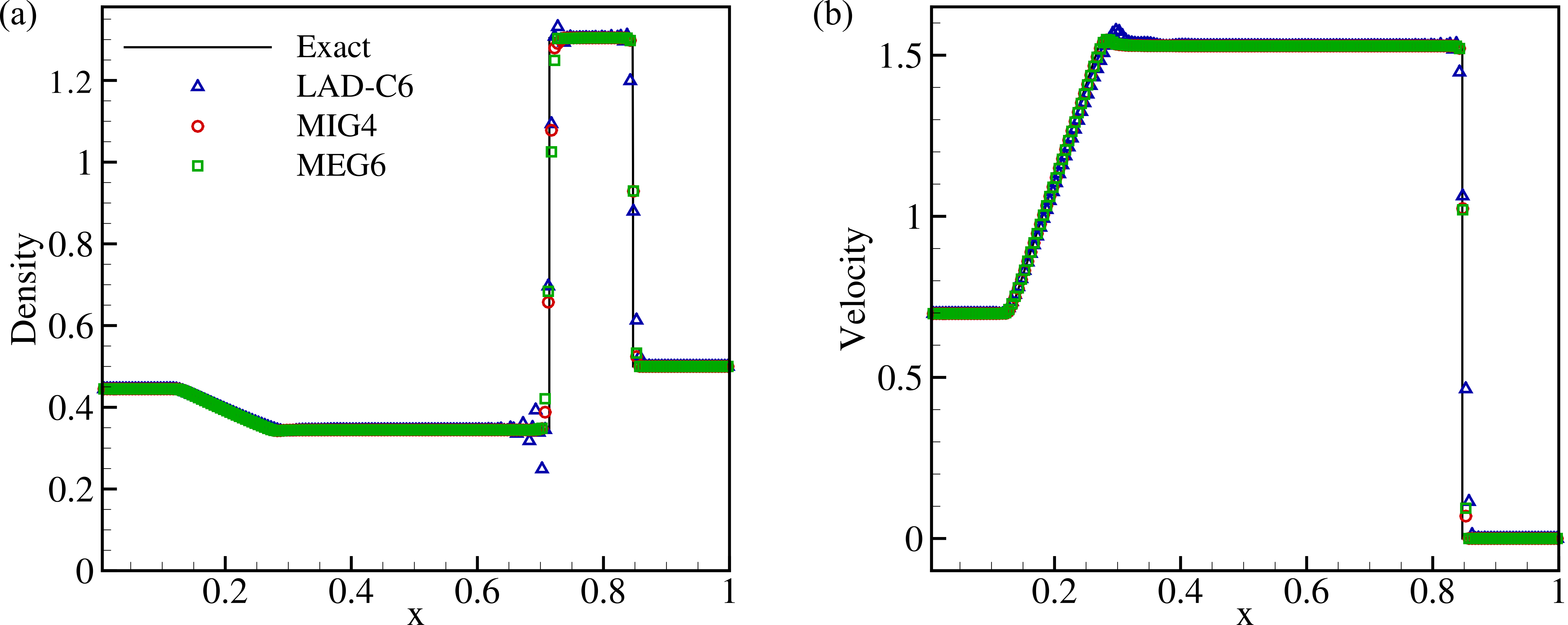}
    \caption{Plots of (a) density and (b) velocity profiles for the Lax problem compared against the exact solution.}
    \label{laxp}
\end{figure}

Figure \ref{sodss} (a) and (b) compare the computed density and velocity plots for the Sod shock tube problem against the exact solution. Both the MIG4 and MEG6 schemes accurately capture the rarefaction wave, contact discontinuity and shock wave. The LAD-C6 scheme performs likewise but suffers from oscillations in the presence of contact discontinuities. Figure \ref{laxp} (a) and (b) compare the computed density and velocity plots for the Lax problem against the exact solution. While the LAD-C6 scheme exhibits oscillations at contact discontinuities, MIG4 and MEG6 capture them with improved accuracy. For both test cases, MEG6 and MIG4 capture the shock discontinuity with fewer cells as compared with the LAD-C6 scheme.

The third shock tube test is the Le Blanc problem \cite{loubere2005subcell}, characterised by extreme jumps in density and pressure of the order $10^3$ and $10^9$ respectively. The simulation is carried out in a computational domain of $x=[0,9]$ using N = 900 uniformly spaced grid points. The test case is simulated with a specific heat ratio of $\gamma = 5/3$, until $t = 6$ units. The initial conditions are given as 

\begin{equation}
(\rho, u, p) =
\begin{cases}
(1.0,\ 0,\ \dfrac{2}{3} \times 10^{-1}), & 0 < x < 3.0, \\
(10^{-3},\ 0,\ \dfrac{2}{3} \times 10^{-10}), & 3.0 < x < 9,
\end{cases}
\end{equation}

\begin{figure}
    \centering
    \includegraphics{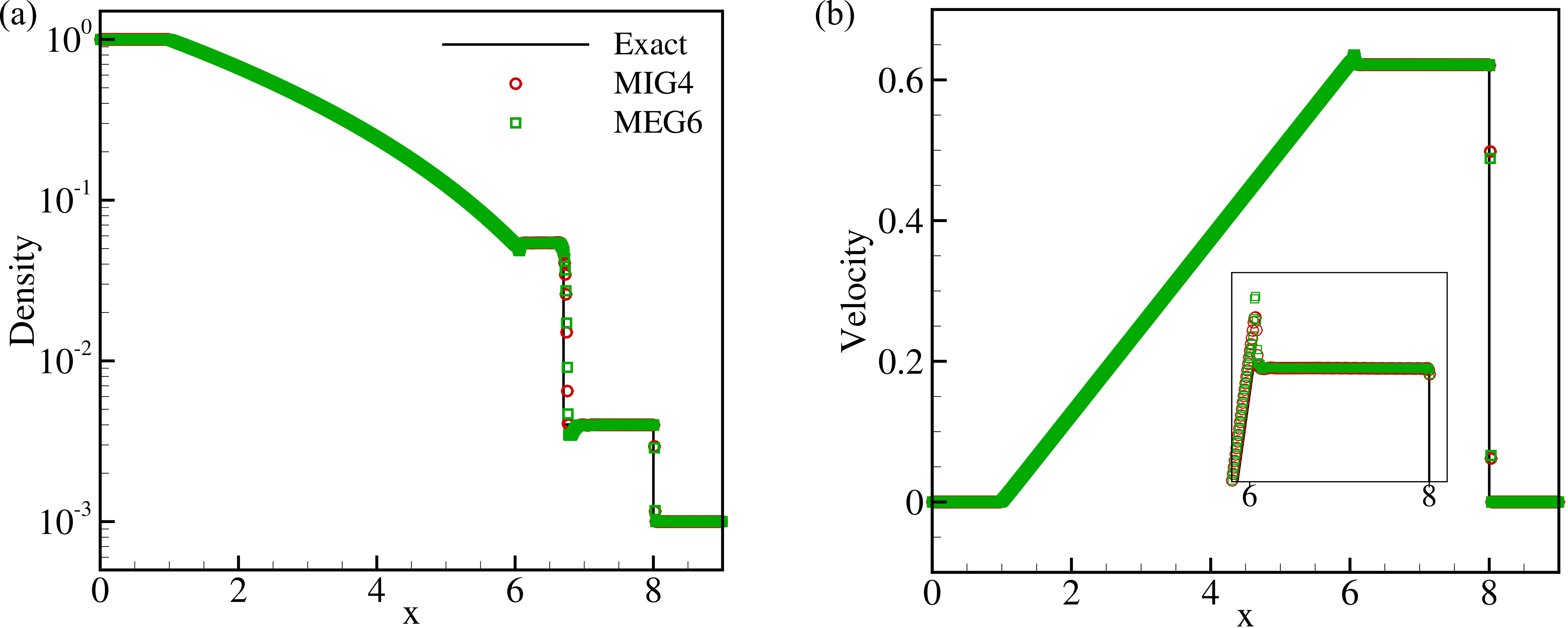}
    \caption{Plots of (a) density and (b) velocity profiles for the Le Blanc problem compared against the exact solution.}
    \label{leb}
\end{figure}

The inflow boundary condition is set to the initial conditions, while variables are extrapolated at the outflow boundary. For this case, the LAD-C6 scheme diverged near the initial discontinuity during the early iterations and failed to produce a stable solution, unlike the C-GBR simulations, which were stable despite the sharp discontinuities. Figures \ref{leb}(a,b) show that the predictions of the MIG4 and MEG6 schemes compare favorably against the exact solution. Although oscillations in velocity are evident while resolving the contact discontinuity, these are marginal when compared with the TENO5 schemes as reported in the literature \cite{chamarthi2023gradient}.

   \begin{figure}
    \centering
    \includegraphics{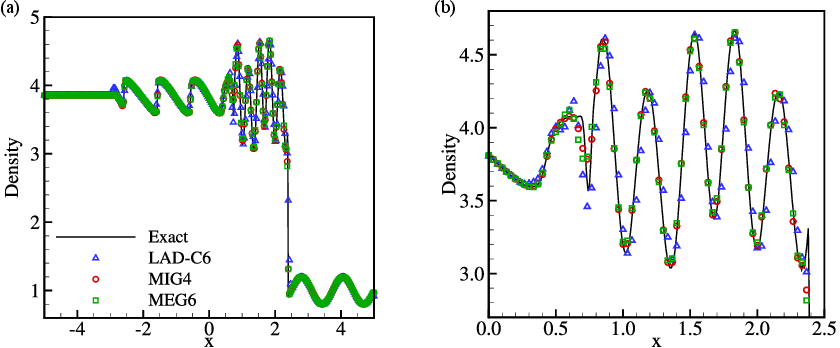}
    \caption{Plots of (a) density profiles and (b) magnified view of the density field ($0 \leq x \leq 2.5$) for the Shu-Osher problem compared against the exact solution \cite{chamarthi2023efficient}.}
    \label{shuo}
\end{figure}

\subsubsection*{Shu-Osher Problem}
A one-dimensional shock–entropy wave interaction problem originally proposed by Shu and Osher \cite{shu1988efficient} is simulated. This test case involves a Mach 3 shock wave interacting with upstream sinusoidal density perturbations, producing both high-frequency oscillations and strong discontinuities. This test is useful for evaluating the robustness and resolution capabilities of shock-capturing schemes in the presence of localised high-frequency features. The simulation is carried out on a computational domain of $x=[-5,5]$ using N = 300 uniformly spaced grid points up to a final time of $t = 1.8$ units. The initial conditions are given as

\begin{equation} \label{shu-osher}
(\rho, u, p) =
\begin{cases}
(3.857, 2.629, 10.333), & \text{if } -5 \leq x < -4, \\
(1 + 0.2 \sin(5(x - 5)), 0, 1), & \text{if } -4 \leq x \leq 5.
\end{cases}
\end{equation}

The inflow boundary condition is set to the initial conditions, while variables are extrapolated at the outflow boundary.
Figure \ref{shuo}(a) compares the computed density for the Shu-Osher problem against the exact solution \cite{chamarthi2023efficient}. A magnified view of the density field between $0<x<2.5$ is shown in Figure \ref{shuo}(b). All schemes perform equally well in capturing the high-frequency density amplitudes, with slight numerical oscillations observed at $x=0.7$ using the LAD-C6 scheme. 

\subsection{Two-dimensional inviscid test case}

\subsubsection*{Shock Entropy Wave Test}

A two-dimensional shock entropy wave test \cite{deng2019fifth} simulation is carried out as an extension of the one-dimensional test case described by Equation \ref{shu-osher}. The computational domain spans $x \times y  = [-5, 5] \times [-1, 1]$ with a uniform grid of $N_x \times N_y = 400 \times 80$. The initial conditions ($\theta= \pi/6$) for this test case are given as

\begin{equation}
(\rho, u, p) =
\begin{cases}
(3.857143,\ 2.629369,\ 10.3333), & x < -4, \\
(1 + 0.2 \sin(10x \cos\theta + 10y \sin\theta),\ 0,\ 1), & \textit{otherwise}.
\end{cases}
\end{equation}

\begin{figure}
    \centering
    \includegraphics{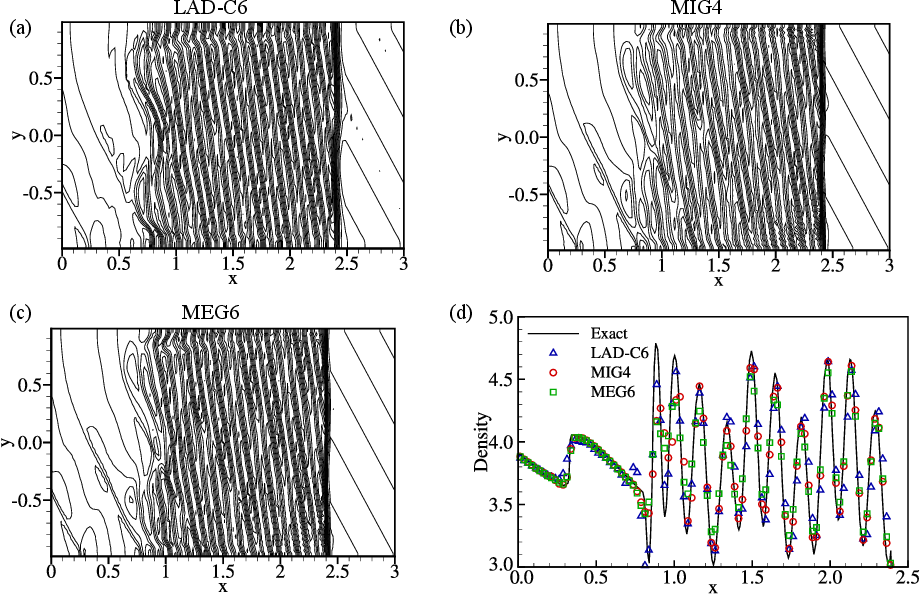}
    \caption{Line contours of density for the 2D shock entropy wave test with (a) LAD-C6, (b) MIG4 and (c) MEG6 schemes; (d) Density variation along the streamwise direction at $y=0$ obtained using various schemes compared against the exact solution \cite{chamarthi2021high}.}
    \label{shentr}
\end{figure}

An inflow condition (same as the initialisation) is used at the left boundary, while the variables are extrapolated at the right boundary. Periodicity is imposed on the top and bottom walls. The simulation is performed until $t = 1.8$ units. Figure \ref{shentr} (a)-(c) shows the density contours flooded by lines obtained with various schemes. The flow features are accurately resolved by the MIG4 scheme as compared with the MEG6 and LAD-C6 schemes. Moreover, the MIG4 and MEG6 schemes exhibit a noticeably sharper shock resolution. Figure \ref{shentr} (d) compares the density contours at $y=0$ obtained using various schemes against the exact solution \cite{chamarthi2021high}. All schemes perform equally well in capturing high-frequency density amplitudes for this test case. 

\subsubsection*{Richtmyer–Meshkov Instability}
The Richtmyer-Meshkov instability phenomenon arises when a planar shock wave interacts with a perturbed interface separating two fluids of different densities. A single-mode two-dimensional RMI \cite{terashima2009front} is simulated assuming that both fluids have a specific heat ratio of $\gamma = 1.4$. The computational domain spans $x \times y  = [0, 4] \times [0, 1]$ with a uniform grid of $N_x \times N_y = 320 \times 80$. The initial conditions are given as follows.

\begin{equation}
(\rho, u, v, p) =
\begin{cases}
(5.04, 0, 0, 1), & \text{if } x < 2.9 - 0.1 \sin(2\pi(y + 0.25)),\ \text{perturbed interface}, \\
(1, 0, 0, 1), & \text{if } x < 3.2, \\
(1.4112,\ -665/1556,\ 0,\ 1.628), & \text{otherwise}.
\end{cases}
\end{equation}

\begin{equation} \label{eq:smooth}
\varphi = \varphi_L (1 - f_{sm}) + \varphi_R f_{sm}, \quad
f_{sm} = \tfrac{1}{2} \left(1 + \text{erf} \left( \tfrac{\Delta D}{C_i \sqrt{\Delta x \Delta y}} \right) \right) 
\end{equation}

\begin{figure}
    \centering
    \includegraphics{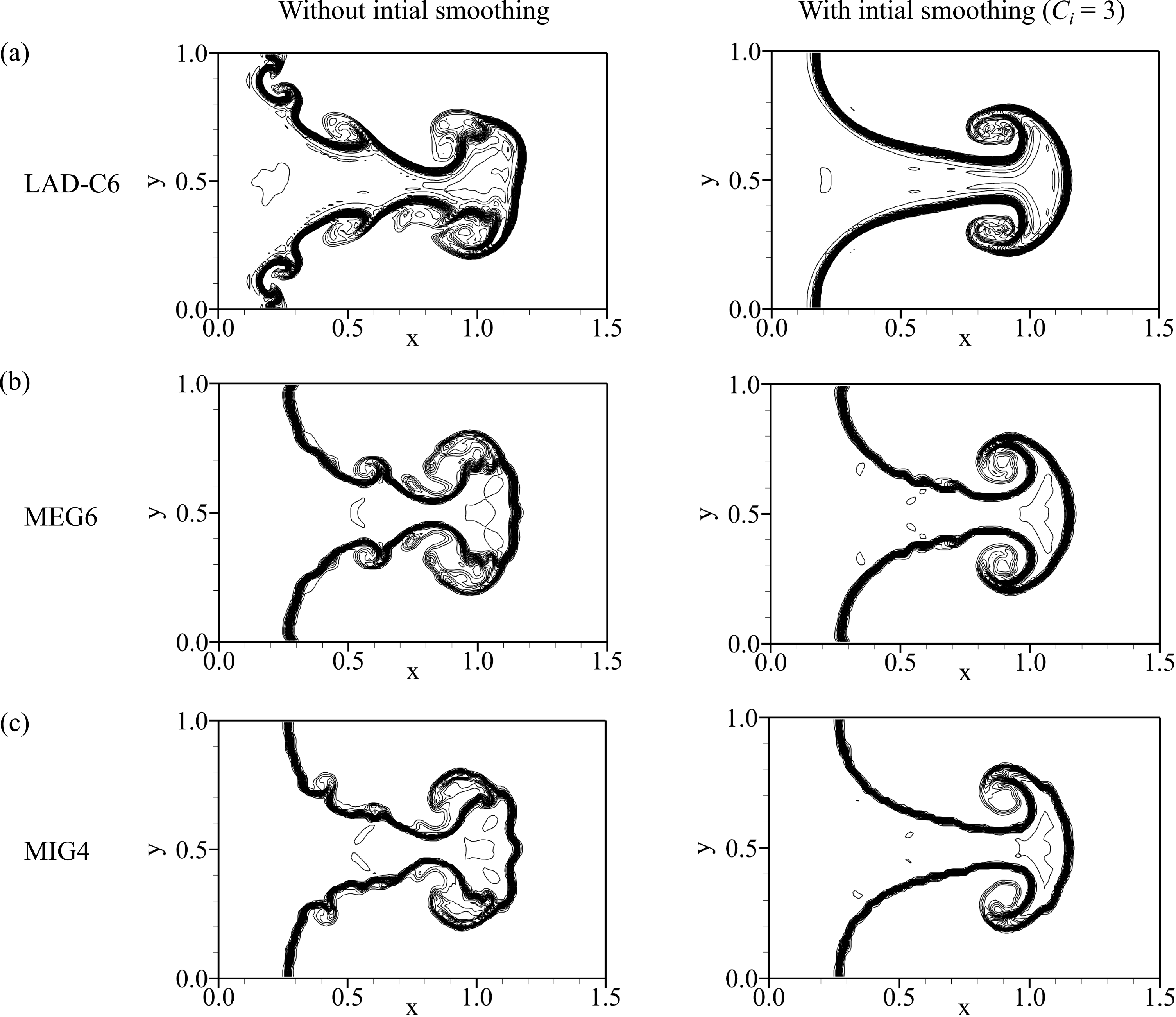}
    \caption{Density contour of the Richtmyer–Meshkov instability with 20 uniformly spaced contour levels with (a) MEG6, (b) MIG4, and (c) LAD-C6 scheme with and without initial smoothing of initial perturbations.}
    \label{richt}
\end{figure}

The left and right boundaries are maintained at initial conditions, while the variables are extrapolated at the outflow. Periodicity is imposed on the top and bottom walls ,and the simulation is performed until $t = 9$ units. Two sets of simulations are performed, with and without the smoothing of the initial sharp interface. Following \textcolor{black}{Wong and Lele}
 \cite{wong2017high}, the primitive variables ($\varphi = (\rho,u,p)$) across the initial sinusoidal interface are smoothed using equation \ref{eq:smooth} with a smoothing parameter $C_i = 3$. Figures \ref{richt} (a)-(c) show the density contours flooded by lines obtained with different schemes without initial smoothing (left) and with initial smoothing (right). In the absence of initial smoothing, both the MEG6 and MIG4 schemes capture small-scale roll-up vortices. However, the results from LAD-C6 schemes show a noticeable asymmetry and distortion of the roll-up vortices. The issue can be attributed to the development of artificial secondary instabilities on a Cartesian grid due to a sinusoidal sharp interface. Following initial smoothing, all of the schemes yield comparable results, although the small-scale vortex structures are not evident. The material interface appears thinner with the MIG4 scheme, indicating a lower numerical dissipation.

\subsubsection*{Shock Vortex Interaction}
A two-dimensional inviscid shock-vortex interaction test case is considered based on the computational setup of Rault et al. \cite{rault2003shock}. The simulation domain spans $x \times y  = [0, 2] \times [0, 1]$ with a uniform grid of $N_x \times N_y = 1024 \times 512$. A stationary normal shock is located at $x=0.5$, and the vortex is placed at the $(x_c, y_c) = (0.25, 0.5)$. The conditions upstream of the shock, excluding those within the vortical region, are specified as $\rho_u = 1$, $u_u = M_s \sqrt{\gamma}$, $v_u = 0$, and $p_u = 1$ where $M_s = 1.1$ is the Mach number upstream of the normal shock. The post-shock conditions can be found using the upstream values of the flow and the normal-shock relations. A vortex of strength $M_v = 1.7$, defined as the ratio of $M_v = v_m/ \sqrt{\gamma}$, is superimposed on the flow upstream of the shock. The angular velocity of the vortex is given as

\begin{equation}
{v}(r) = v_\theta(r) + u_\infty \, \hat{e}_x, \qquad
v_\theta(r) = \hat{e}_{\theta}.
\begin{cases}
v_m \dfrac{r}{a} & \text{if } r \leq a, \\
v_m \dfrac{a}{a^2 - b^2} \left( r - \dfrac{b^2}{r} \right) & \text{if } a \leq r \leq b, \\
0 & \text{if } r > b,
\end{cases}
\end{equation}

\begin{figure}
    \centering
    \includegraphics{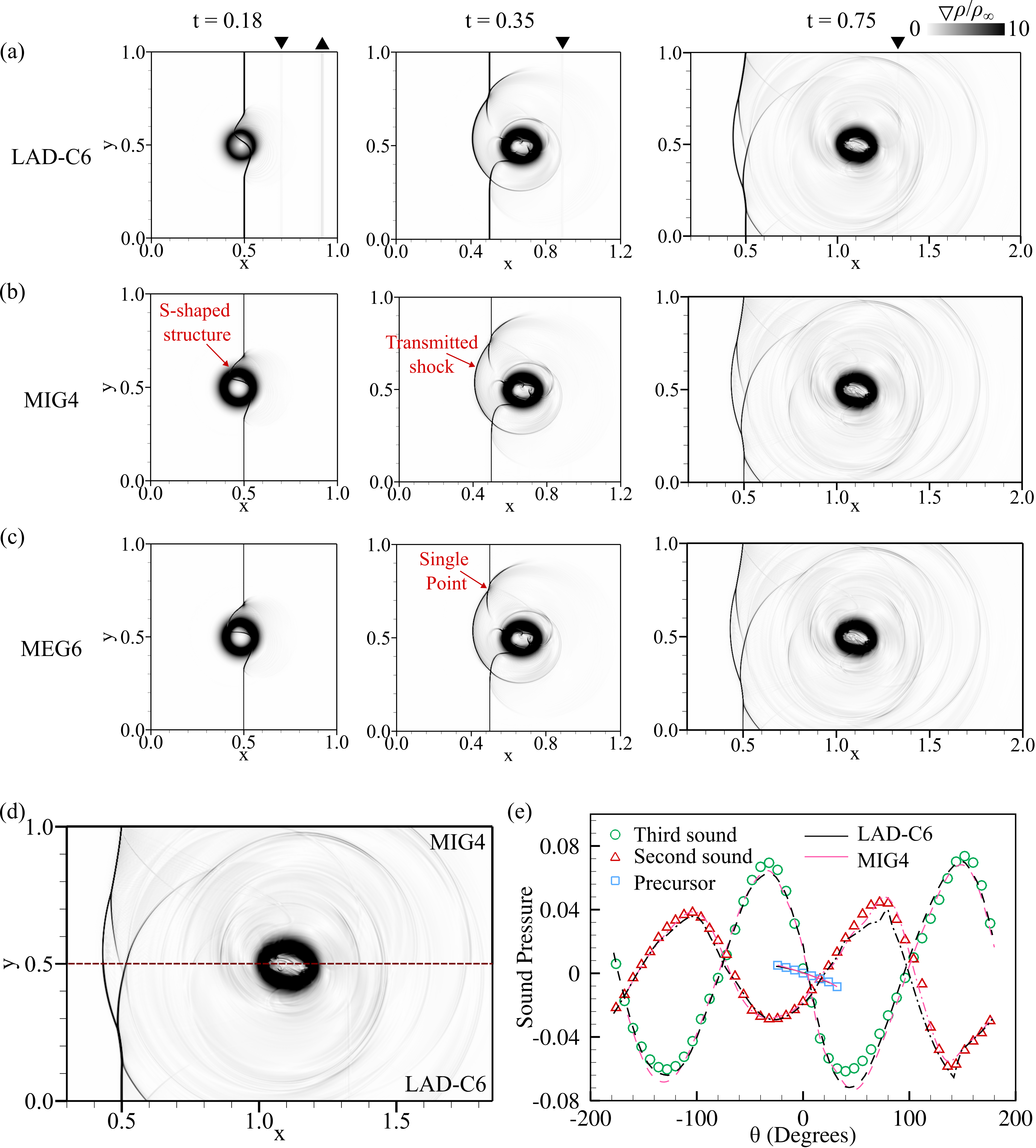}
    \caption{Contours of density gradient magnitude computed using (a) LAD-C6, (b) MIG4, and (c) MEG6 at time intervals of $t = 0.18,  0.35\ \text{and}\ 0.75$ units; (d) Top: MIG4 and bottom: MEG6 at $t = 0.75$ units; (e) Plots of predicted sound pressure along the three acoustic waves compared with the reference results of Rault et al. \cite{rault2003shock}.}
    \label{shvort}
\end{figure}

where $\hat{e}_x$ and $\hat{e}_{\theta}$ are the unit vectors along the streamwise and tangential directions, respectively, and $v_m$ is the maximal angular velocity. The radius of the inner core is specified as $a=0.075$ and the radius of the outer core is $b=0.175$. As noted by Fernandez et al. \cite{fernandez2018posteriori}, the vortex can be interpreted as a perturbation imposed on the previously determined upstream flow values. The resultant velocity within the vortical region can then be specified as 

\begin{equation}
u_{\text{vor}}(r) = u_u - \dfrac{v_m r}{a} \dfrac{y - y_c}{r}, \qquad
v_{\text{vor}}(r) = v_u + \dfrac{v_m r}{a} \dfrac{x - x_c}{r}
\end{equation}

where $u_u$ and $v_u$ are the velocity upstream of the normal shock. The thermodynamic variables such as temperature, density, and pressure within the vortex core can be found using the following expressions. 

\begin{equation}
\dfrac{dT_{\text{vor}}(r)}{dr} = \dfrac{\gamma - 1}{R \gamma} \dfrac{v_\theta^2(r)}{r}, \qquad
\rho_{\text{vor}}(r) = \rho_u \left( \dfrac{T_{\text{vor}}(r)}{T_u} \right)^{\dfrac{1}{\gamma - 1}}, \qquad
p_{\text{vor}}(r) = p_u \left( \dfrac{T_{\text{vor}}(r)}{T_u} \right)^{\dfrac{\gamma}{\gamma - 1}}
\end{equation}

A supersonic inflow boundary condition is imposed at the inlet, while a subsonic outflow condition (static pressure downstream of the shock is specified and other variables are extrapolated) is applied at the outlet. The simulation is carried out until $t = 0.75$. Figures \ref{shvort} (a)-(c) illustrate the temporal evolution of density gradient magnitude contours obtained using different schemes at time intervals of $t = 0.18,  0.35\ \text{and}\ 0.75$ units. The strong vortex convects downstream and interacts with the weak shock. As the vortex core passes through the normal shock ($t=0.18$), the latter distorts to an S-shaped structure around the vortex. At $t=0.35$, the transmitted shock wave moves beyond the vortex field. As noted by Rault et al. \cite{rault2003shock}, the shock strength is weaker than the vortex strength, indicating a regular reflection phenomenon in secondary shock structures that join up with the primary shock at a single point. Both the MEG6 and MIG4 schemes demonstrate a sharper resolution of flow structures compared to the LAD-C6 scheme. Moreover, the density gradient contour of the LAD-C6 scheme exhibits a weak (down triangle symbol) and a strong (up triangle symbol) wavelike disturbance which propagates downstream and gradually diminishes. Figure \ref{shvort} (d) compares the density gradient contours of the MIG4 and LAD-C6 schemes. The MIG4 scheme captures the shock within a few cells compared to the LAD-C6 scheme, indicating low dissipation properties of the former scheme. The shock-vortex interaction results in the formation and evolution of strong acoustic waves, namely, the precursor, second, and third sound waves. Figure \ref{shvort} (e) plots the circumferential distribution of the sound pressure $\Delta P = (P-P_2)/P_2$ at three different radii of $r=0.85,\ 0.48 \ \text{and}\ 0.22$, where $P$ is the local pressure and $P_2$ is the reference pressure downstream of the shock. For both schemes, the results are in favourable agreement with those of Rault et al. \cite{rault2003shock}.

\subsubsection*{Kelvin Helmholtz instability}
A Kelvin–Helmholtz instability (KHI) is simulated, where velocity gradients at the initial interface result in the formation of vortices and distinct mixing patterns. The KHI is simulated over a periodic domain of $x \times y  = [0, 1] \times [0, 1]$ on a grid comprising $N_x \times N_y = 512 \times 512$ until a final time of $t = 0.8$. The initial conditions are given as follows.

\begin{equation}
\begin{aligned}
    &p(x,y) = 2.5, \quad 
    \rho(x, y) = 
    \begin{cases}
        2, & \text{if } 0.25 < y \leq 0.75, \\
        1, & \text{else},
    \end{cases} \\ 
    &u(x, y) = 
    \begin{cases}
        0.5, & \text{if } 0.25 < y \leq 0.75, \\
        -0.5, & \text{else},
    \end{cases} \\ 
    &v(x, y) = 0.1 \sin(4\pi x) 
    \left\{
    \exp\left[-\frac{(y - 0.75)^2}{2\sigma^2}\right] +
    \exp\left[-\frac{(y - 0.25)^2}{2\sigma^2}\right]
    \right\}, \quad \text{where } \sigma = \frac{0.05}{\sqrt{2}}.
\end{aligned}
\end{equation}

\begin{figure}
    \centering
    \includegraphics{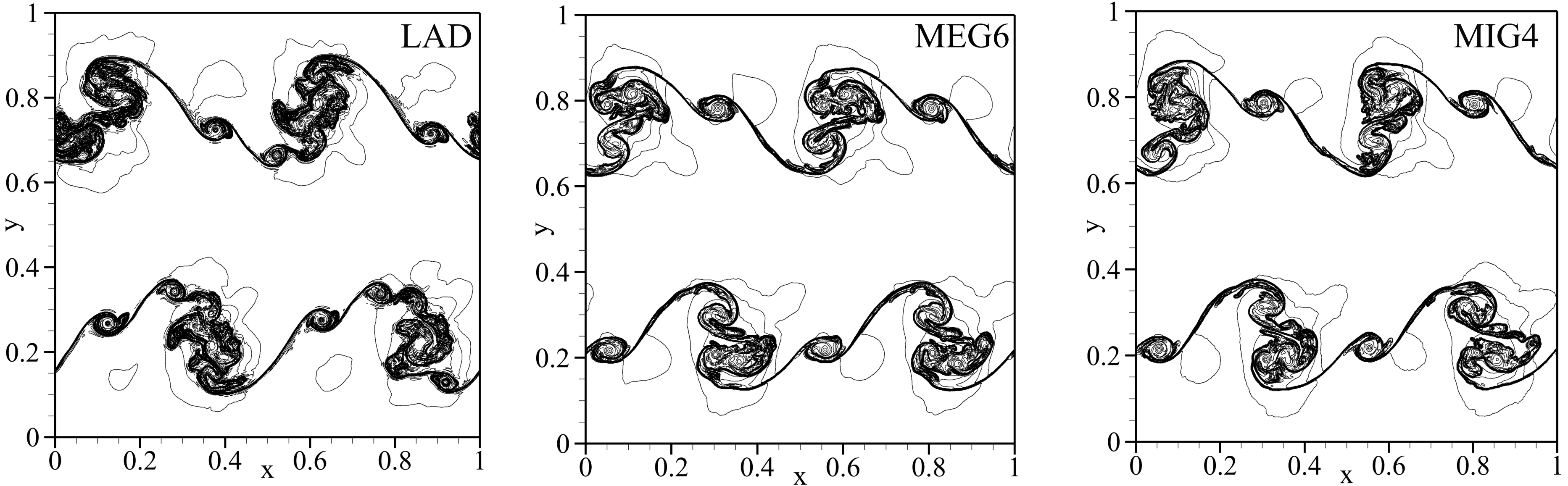}
    \caption{Density contour of the Kelvin Helmholtz instability test case with 20 uniformly spaced contour levels simulated using (a) LAD-C6, (b) MEG6, and (b) MIG4 schemes.}
    \label{khi}
\end{figure}

Figures \ref{khi} (a)-(c) show the line density contours of the Kelvin-Helmholtz instability simulated using different schemes. All schemes capture small-scale vortices and complex flow features. As observed in the previous test cases, both MEG6 and MIG4 resolve the contact discontinuity free of oscillations and with a thinner material interface. However, high-frequency oscillations are evident across the interface captured using the LAD-C6 scheme.

\subsubsection*{2D Riemann Problem}
Two Riemann problems, one strong and one weak, are chosen from Shulz-Rinne et al. \cite {schulz1993numerical} to evaluate the performance of different schemes. The 2D Riemann problem is initialised in a square domain of $  x\times y = [0, 1] \times [0, 1]$, where each quadrant is assigned different values of the primitive variables. The interaction of these discontinuities generates slip lines along which Kelvin–Helmholtz instabilities develop, forming small-scale structures. This test case serves as a benchmark for assessing the numerical dissipation of a given scheme. The initial conditions for the strong Riemann test case (configuration 3 in \cite {schulz1993numerical}) are given as

\begin{equation}
(\rho, u, v, p) =
\begin{cases}
(1.5,\ 0,\ 0,\ 1.5), & \text{if } x > 0.8,\ y > 0.8, \\
(33/62,\ 4/\sqrt{11},\ 0,\ 0.3), & \text{if } x \leq 0.8,\ y > 0.8, \\
(77/558,\ 4/\sqrt{11},\ 4/\sqrt{11},\ 9/310), & \text{if } x \leq 0.8,\ y \leq 0.8, \\
(33/62,\ 0,\ 4/\sqrt{11},\ 0.3), & \text{if } x > 0.8,\ y \leq 0.8.
\end{cases}
\end{equation}

\begin{figure}
    \centering
    \includegraphics{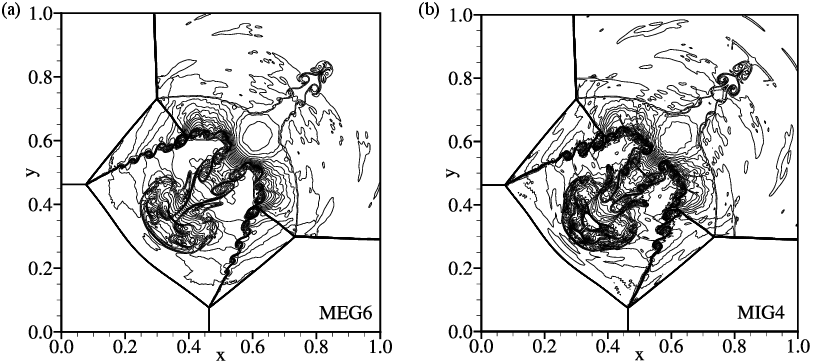}
    \caption{Density contour of the strong 2D Riemann problem with 30 uniformly spaced contour levels with (a) MEG6 and (b) MIG4 scheme.}
    \label{riemann3}
\end{figure}

Extrapolation boundary conditions are imposed on all walls. The test case is simulated on a uniform grid comprising $N_x \times N_y = 400 \times 400$ until a final time of $t = 0.8$ units. Notably, the initial maximum pressure ratio across the quadrants is $\approx 51$, and the LAD-C6 scheme diverged at the intersection of the quadrants in the initial iterations. Alternate strategies to achieve converged solution for this case with LAD are discussed in Section \ref{hybrid}. Figure \ref{riemann3} illustrates density contours of the strong Riemann test case with 30 uniformly spaced contour levels flooded by lines with the MEG6 and MIG4 schemes. Both MEG6 and MIG4 schemes capture vortical and jet-like structures, but the latter scheme produces more pronounced small-scale vortical structures. 

The weak Riemann problem (configuration F in \cite {schulz1993numerical}) is simulated with the following initial condition.

\begin{equation}
(\rho, u, v, p) =
\begin{cases}
(0.5313,\ 0,\ 0,\ 0.4), & \text{if } x > 0.5,\ y > 0.5, \\
(1,\ 0.7276,\ 0,\ 1), & \text{if } x \leq 0.5,\ y > 0.5, \\
(0.8,\ 0,\ 0,\ 1), & \text{if } x \leq 0.5,\ y \leq 0.5, \\
(1,\ 0,\ 0.7276,\ 1), & \text{if } x > 0.5,\ y \leq 0.5.
\end{cases}
\end{equation}

\begin{figure}
    \centering
    \includegraphics{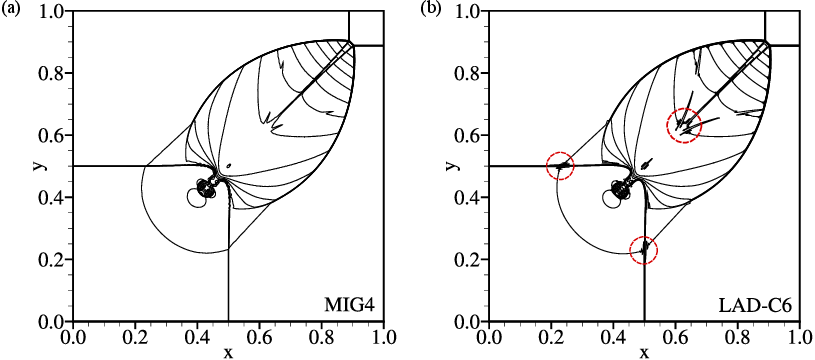}
    \caption{Density contour of the weak 2D Riemann problem with 25 uniformly spaced contour levels with (a) MIG4 and (b) LAD-C6 scheme.}
    \label{riemann12_1}
\end{figure}

\begin{figure}
    \centering
    \includegraphics{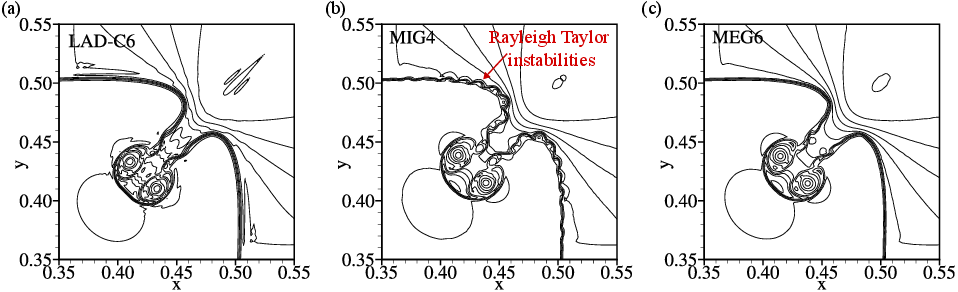}
    \caption{Zoomed in density contour of the weak 2D Riemann problem with 30 equally spaced contour levels with (a) LAD-C6, (b) MIG4, and (c) MEG6 scheme.}
    \label{riemann12_2}
\end{figure}

The test case is simulated on a uniform grid comprising $N_x \times N_y = 800 \times 800$ until a final time of $t = 0.25$ units. The initial maximum pressure ratio across the quadrants is 2.5 which is significantly lower than the strong Riemann case discussed earlier. The simulations using the LAD-C6 scheme are found to be stable for this weak Riemann case. Figure \ref{riemann12_1} illustrates 30 uniformly spaced density contour lines predicted using the MIG4 and LAD-C6 scheme. The results from LAD-C6 schemes show noticeable spurious oscillations at a few locations, as indicated by the red circles in the density contour, while the results of the MIG4 scheme are relatively superior. Figure \ref{riemann12_2} also shows a magnified view of the density contours $  x\times y = [0.35,0.55] \times [0.35,0.55]$ that highlights the jet-like structure where vortices develop in the lower left quadrant. All schemes show a consistent central structure, although Rayleigh-Taylor instabilities appear near the trailing region for the MIG4 scheme. Inline with the previous observations, the results from MIG4 and MEG6 are sharper and free from oscillations compared to the LAD-C6 scheme. Overall, for both the weak and strong 2D Riemann cases, the GRB schemes exhibit low numerical dissipation and are robust in capturing sharp discontinuities and multi-scale features. 

\subsubsection*{Compressible Triple Point Problem} \label{comp3p_case}
A compressible triple point problem is a two-dimensional multi-material Riemann problem with three different initial states and is widely used to assess the robustness and accuracy of multi-phase compressible flow solvers. For the present case, a simplified problem is considered with all the regions modelled as air with a specific heat ratio of $\gamma = 1.4$. The computational domain spans $x \times y  = [0, 7] \times [0, 3]$ discretized with a uniform grid of $N_x \times N_y = 1792 \times 768$. The initial conditions from Pan et al. \cite{pan2018conservative}, assuming all regions as ideal air, are given as

\begin{equation}
(\rho, u, v, p, \gamma) = 
\begin{cases}
(1.0,\ 0,\ 0,\ 1.0,\ 1.4), & \text{sub-domain } [0, 1] \times [0, 3], \\
(1.0,\ 0,\ 0,\ 0.1,\ 1.4), & \text{sub-domain } [1, 7] \times [0, 1.5], \\
(0.125,\ 0,\ 0,\ 0.1,\ 1.4), & \text{sub-domain } [1, 7] \times [1.5, 3].
\end{cases}
\end{equation}

\begin{figure}
    \centering
    \includegraphics{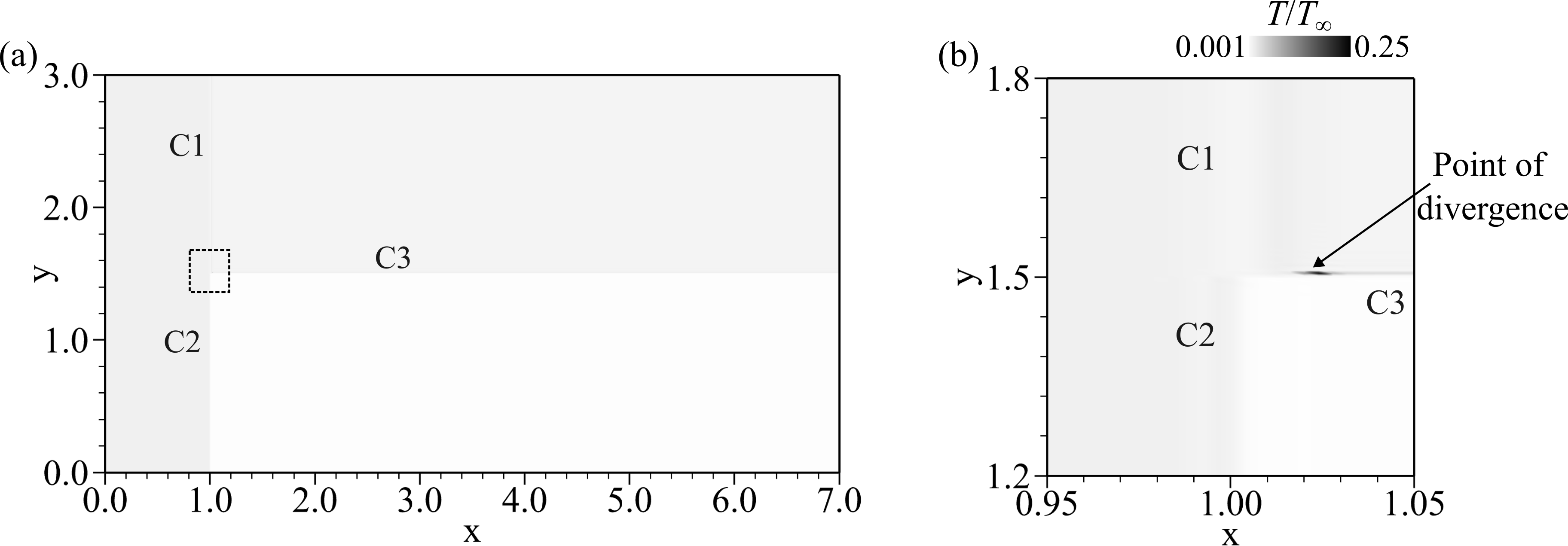}
    \caption{Contours of temperature using LAD-C6 scheme (a) illustrating the flow-field and (b) a zoomed-in view indicating the point of divergence.}
    \label{LAD_comp3p}
\end{figure}

\begin{figure}
    \centering
    \includegraphics{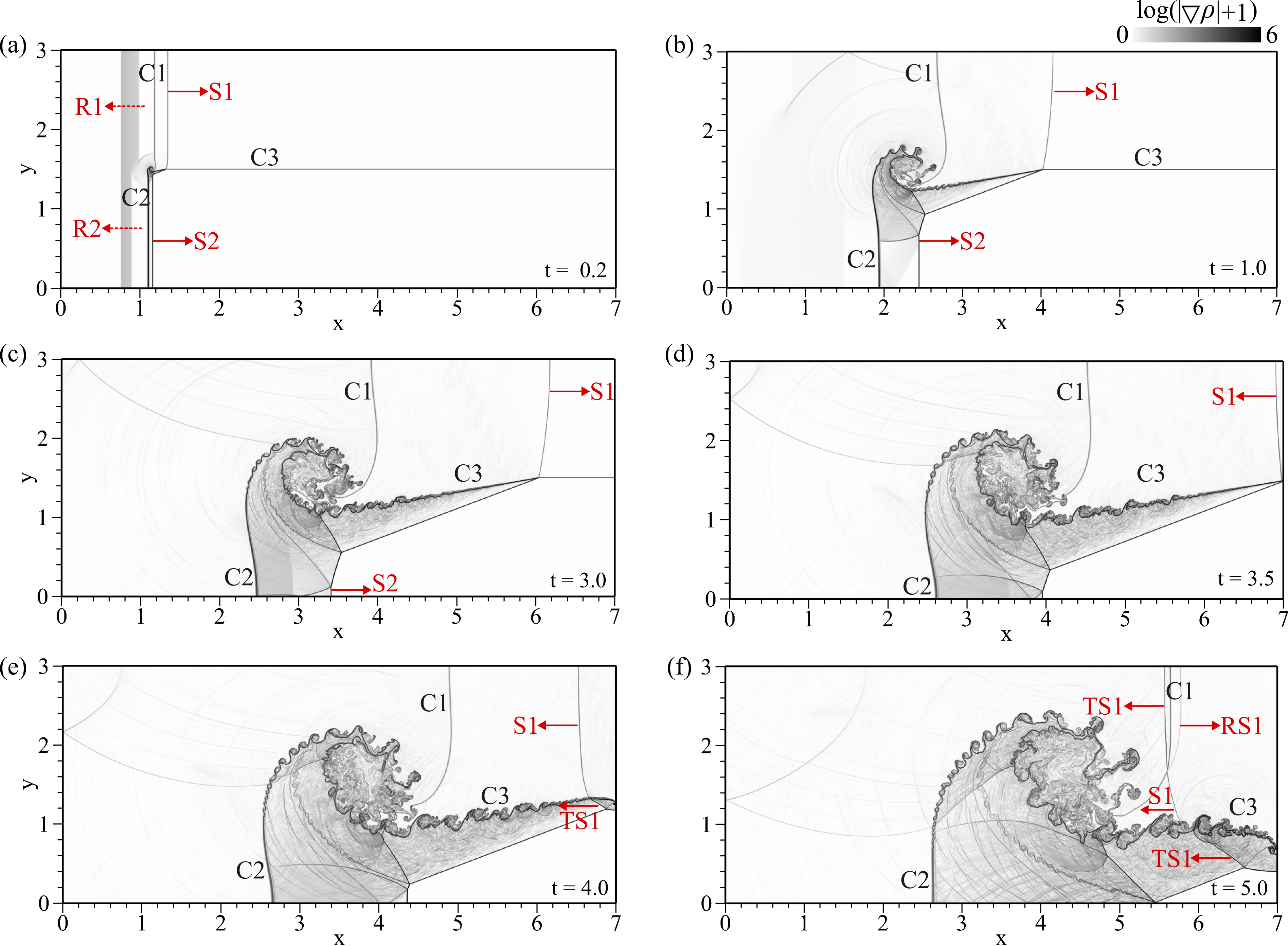}
    \caption{Time evolution of compressible triple point shown through contours of density gradient magnitude using MIG4 scheme at time intervals of (a) $t=0.2$, (b) $t=1.0$, (c) $t=3.0$, (d) $t=3.5$, (e) $t=4.0$, and (f) $t=5.0$}
    \label{comp3p}
\end{figure}

Extrapolation boundary conditions are imposed on all walls, and the test case is simulated for $t=5.0$ units. The LAD-C6 scheme failed to produce a stable simulation. Figures \ref{LAD_comp3p} (a) and (b) illustrate the initial flow field and a magnified region of the problematic cell before the onset of the divergence observed at the interface of the contact discontinuity (C3). The results of the MIG4 and MEG6 schemes are qualitatively similar and with minimal differences; therefore, the results are presented only with the MIG4 scheme. Figure \ref{comp3p} shows the time evolution of the shock system using the density gradient magnitude contours computed with the MIG4 scheme at time intervals of (a) $t=0.2$, (b) $t=1.0$, (c) $t=3.0$, (d) $t=3.5$, (e) $t=4.0$, and (f) $t=5.0$. The simulation results align closely with the computational predictions of Pan et al. \cite{pan2018conservative}, using a $2 \times$ finer mesh of $3584 \times 1536$ as compared to the present setup. As shown in \ref{comp3p} (a), the initial 2D Riemann problem generates a system of waves, namely, contact discontinuities (C1 and C2), a shock wave travelling to the right boundary (S1 and S2), and a rarefaction wave (R1 and R2) travelling towards the left boundary. Figure \ref{comp3p} (b) illustrates the formation of a roll-up region near the triple point as the shock wave S1 travels faster than S2. The wave interaction in this region creates a complex pattern. The initial rarefaction waves, R1 and R2, gradually weaken and dissipate. The interaction of contact discontinuities C1, C2, and C3 results in the formation of small-scale vortices driven by Kelvin–Helmholtz instabilities, as shown in Figure \ref{comp3p} (c). With the shock wave S1 moving toward the right boundary, it reflects back into the domain and interacts with C3, resulting in the formation of the transmitted shock wave TS1 (Fig. \ref{comp3p} (d,e)). As TS1 moves inward, it interacts with C1, generating a transmitted shock (TS2) and a reflected shock (RS1). It is apparent that all of these complex interactions are well captured by the C-GBR schemes. Although LAD diverged for this case, we have re-attempted this case using a hybrid solver that will be discussed in Section \ref{hybrid}.

\subsubsection*{Shock-Bubble interaction}
A two-dimensional single-species inviscid shock-bubble interaction problem is considered. This test case has been widely studied in the literature \cite{haas1987interaction, wong2017high, chamarthi2023gradient} as a multi-species problem where a helium bubble interacts with the planar shock wave and the surrounding air. The test case is simulated on a computational domain of $  x/D\times y/D  = [0, 6.5] \times [0, 1.78]$ where $D$ is the diameter of the bubble. A planar normal shock of Mach 1.22 is located at $x/D = 4.5$ and the bubble is positioned at $(x_c, y_c) = [3.5D, 0.89D]$. The initial conditions are given as follows.

\begin{equation}
(\rho, u, v, p, \gamma) =
\begin{cases}
(1.3764,\ -0.3336,\ 0,\ 0,\ 1.5698/1.4,\ 1.4), & \text{for post-shock air,} \\
(0.1819,\ 0,\ 0,\ 1/1.4,\ 1.4), & \text{for air cylinder,} \\
(1,\ 0,\ 0,\ 1/1.4,\ 1.4), & \text{for pre-shock air.}
\end{cases}
\end{equation}

\begin{figure}
    \centering
    \includegraphics{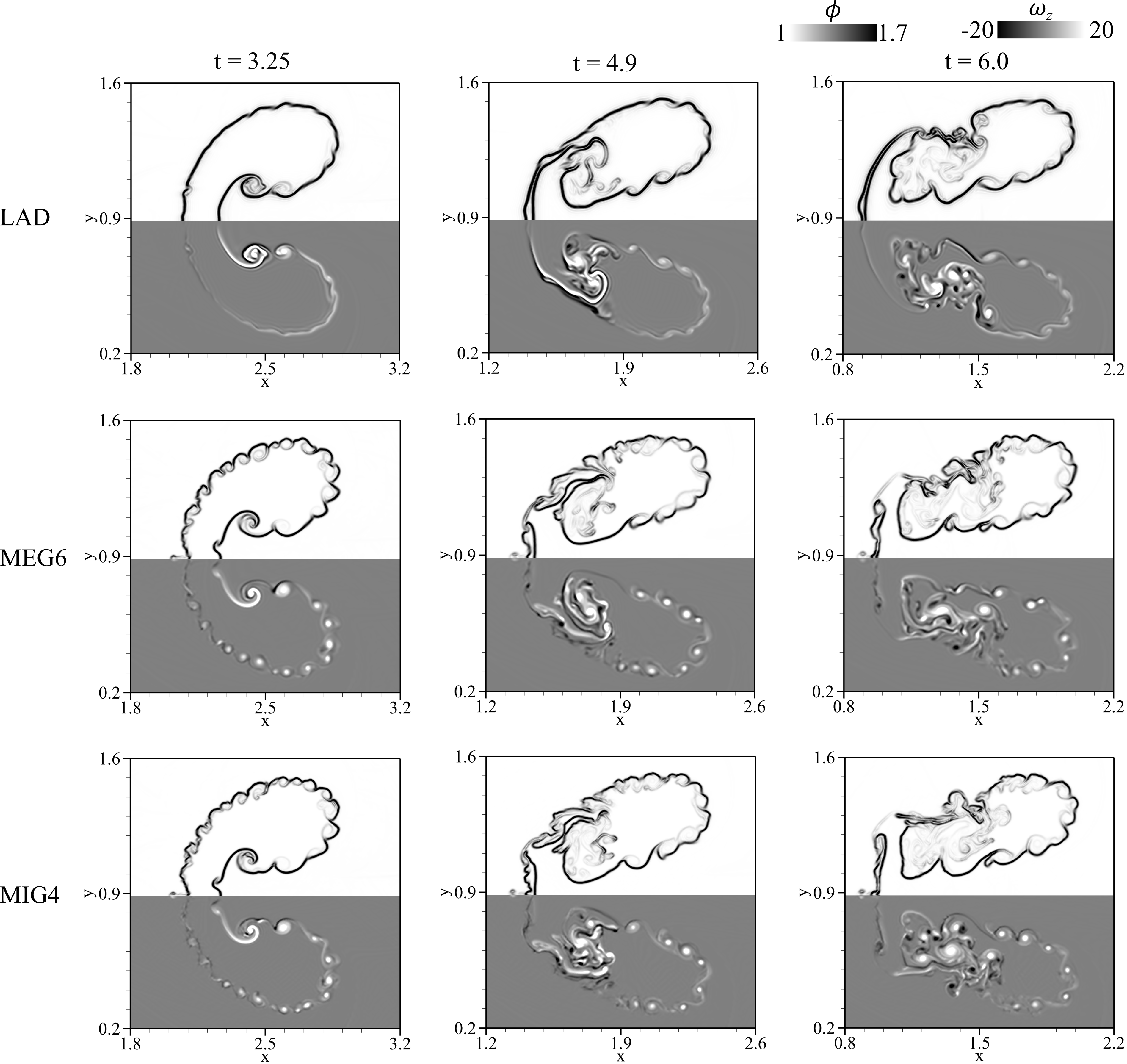}
    \caption{Time evolution of shock-bubble interaction at different time intervals of $t=3.25, 4.9 \ \text{and}\ 6.0$ units computed with (a) LAD-C6, (b) MEG6, and (c) MIG4 schemes. The top and bottom portions of each frame illustrate the normalised density gradient magnitude ($\phi$) and $z$-vorticity ($\omega_z$), respectively, on a computational domain measuring  $1.4D \times 1.4D$.}
    \label{shock_bubble}
\end{figure}

The simulation is carried out on a uniform grid comprising $N_x \times N_y = 1300 \times 356$ until $t=6$ units. Both the left and right boundaries are set as outflows. An inviscid wall boundary condition is imposed on the top and bottom walls. Figure \ref{shock_bubble} illustrates the time evolution of the shock-bubble interaction at different time intervals of $t=3.25, 4.9 \ \text{and}\ 6.0$ units predicted using the LAD-C6, MEG6, and MIG4 schemes. The upper half of the frame shows a non-linear function of density gradients, $\phi = \exp ( |\Delta \rho|/|\Delta \rho|_{\max})$, while the lower half shows the $z$-vorticity ($\omega_z$), on a computational domain measuring $1.4D \times 1.4D$. 

During the initial transients, the shock travels upstream and interacts rapidly with the air bubble rather than the surrounding air. This is attributed to the different sound speeds (due to different densities) within the bubble and the surroundings. Subsequently, the bubble interface flattens and forms a jet from the baroclinic torque that pierces the bubble. The Kelvin–Helmholtz instability induces vortex formation along the bubble interface. The ability to capture these structures is strongly influenced by the grid resolution and the dissipative nature of the numerical scheme. Figure \ref{shock_bubble} shows that the results from all the schemes are stable and non-oscillatory. It is apparent that the LAD-C6 scheme is relatively more dissipative, while the MEG6 and MIG4 schemes accurately resolve the secondary instabilities that lead to fine-scale vortical structures across the material interfaces.

For the 1D and 2D-inviscid test cases examined so far, it is observed that LAD struggles to provide a stable simulation for the test cases with strong discontinuities. These include the 1D Le Blanc, 2D strong Riemann, and 2D compressible triple point cases. In contrast, the C-GBR schemes demonstrate consistent performance and produce a stable solution with minimal numerical dissipation. In the next section, we explore the efficacy of these schemes when applied to viscous test cases.

\subsection{Two-dimensional viscous test cases}

\subsubsection*{Double Periodic Shear Layer}
\begin{figure}
    \centering
    \includegraphics{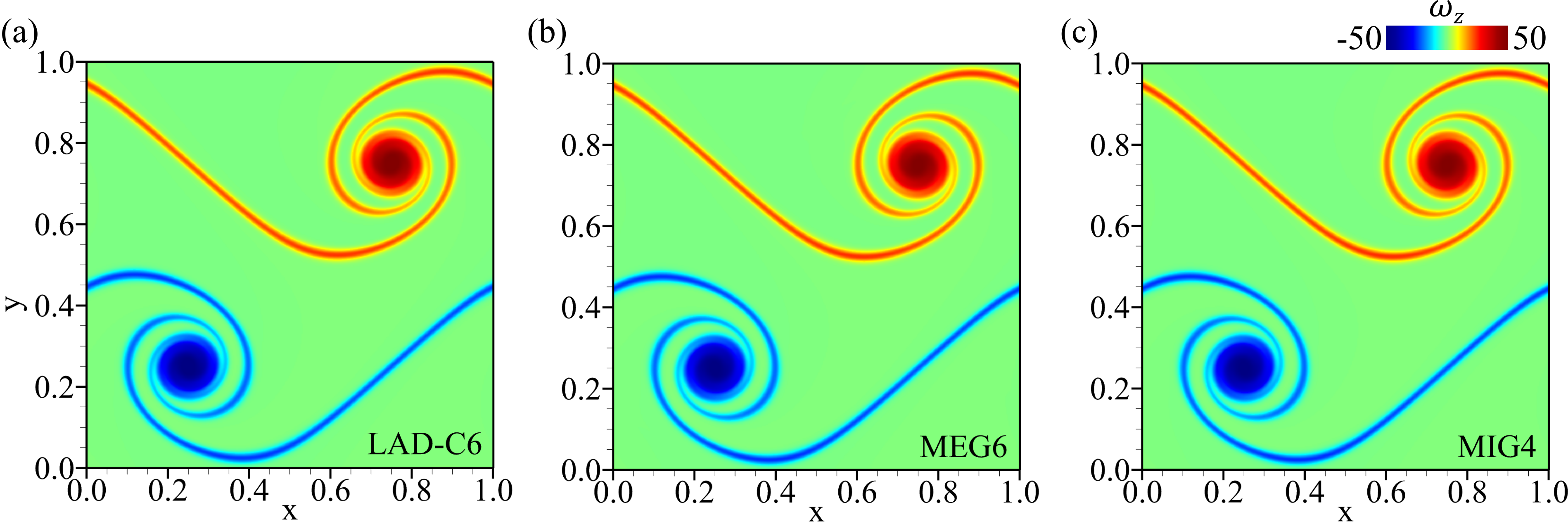}
    \caption{Contours of vorticity ($\omega_z$) at $t=1$ unit obtained with (a) LAD-C6, (b) MEG6, and (c) MIG4 scheme.}
    \label{dpsl1}
\end{figure}

\begin{figure}
    \centering
    \includegraphics{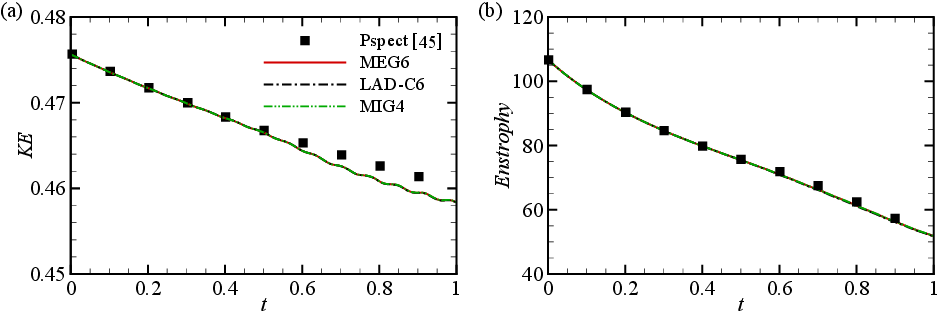}
    \caption{Plots of time evolution of (a) Kinetic energy (KE) and (b) Enstrophy computed using LAD-C6, MEG6 and MIG4 schemes and compared with the pseudospectral (Pspect) results from Minion \& Brown \cite{minion1997performance}.}
    \label{dpsl2}
\end{figure}

A two-dimensional viscous double periodic shear layer (DPSL) test case is simulated on a periodic domain of $  x\times y  = [0, 1] \times [0, 1]$. The Mach number is taken as $M = 0.1$ and the reference Reynolds number is fixed at $Re=10^4$. All the length scales are normalised with a reference length, $L$. The simulation is performed on a uniform grid comprising $N_x \times N_y = 320 \times 320$ until a final time of $t = 1.0$ units. The initial conditions are given as 

\begin{equation}
\begin{aligned}
u(x, y) &=
\begin{cases}
\tanh (\delta_w (y - 0.25)), & \text{if } y \leq 0.5 \\
\tanh (\delta_w (0.75 - y)), & \text{if } y > 0.5
\end{cases} \\
v(x, y) &= \delta_P \sin\left( 2\pi(x + 0.25) \right)
\end{aligned}
\end{equation}

where, the shear layer width is $\delta_w=80$ and $\delta_p=0.05$ is the initial perturbation strength. Figure \ref{dpsl1} compares the $z$-vorticity ($\omega_z$) contours obtained with various schemes after the $t=1$ unit. As observed, the initial two parallel shear layers evolve into dominant vortices. The results from the LAD-C6, MEG6 and MIG4 schemes are qualitatively similar, with no spurious oscillations along the shear layer indicating a physically accurate solution. Figure \ref{dpsl2} further compares the temporal evolution of the kinetic energy (KE) and enstrophy calculated using different schemes against the pseudospectral (Pspect) results from Minion \& Brown \cite{minion1997performance}. All the schemes compare favourably and agree well with the reference data.

\subsubsection*{2D Laminar Hypersonic Compression Ramp}
An interaction of a two-dimensional laminar shock boundary layer on a compression ramp is simulated. The computational setup is based on the numerical studies by Simeonides et al. \cite{simeonides1994experimental}. The free-stream Mach number is $M_{\infty}$ = 6.0, the Reynolds number is $Re_x=10 \times 10^6/m$, and the compression corner angle is $\theta _c$ = 7.5$^{\circ}$. The distance from the leading edge to the compression corner is 0.04$L_\text{r}$, and the wall-normal extent of the computational domain is $L_{y}$ = 0.04$L_\text{r}$,  where $L_\text{r}=1.0\ m$ is the reference length scale. The computational domain is discretised with grid points $ N_x\times N_y$ = $378 \times 81$ in the streamwise and wall-normal direction, respectively. A uniform inflow boundary condition is specified at the inlet, whereas a supersonic outflow is prescribed at the outlet. A no-slip adiabatic wall condition is imposed on the bottom wall with variables extrapolated at the top wall.

\begin{figure}
    \centering
    \includegraphics{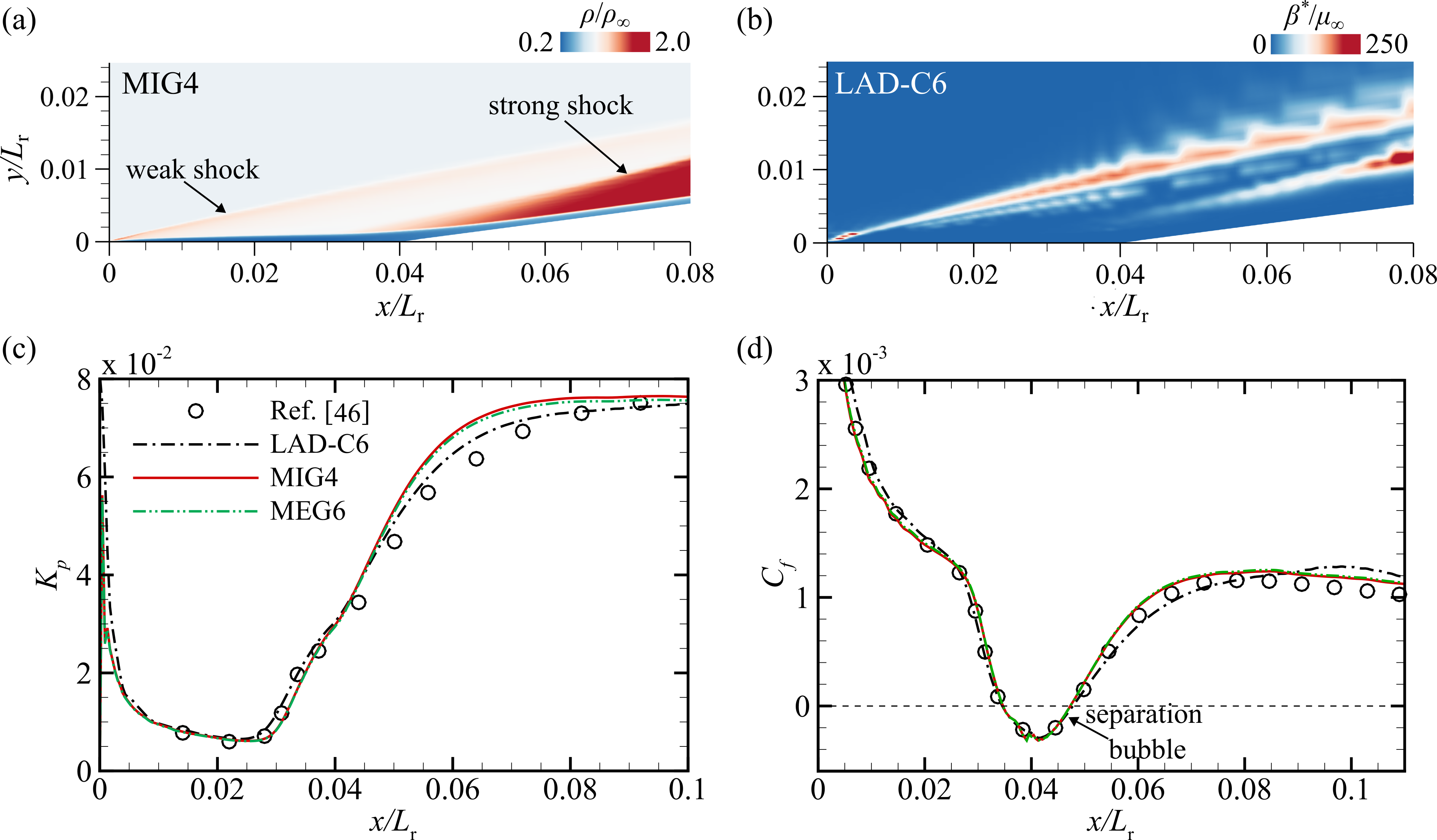}
    \caption{Contours of (a) density with MIG4 scheme and (b)  artificial bulk viscosity ($\beta^*$) across the shock with LAD-C6 scheme; streamwise evolution of (c) wall-pressure coefficient ($K_p$) and (d) skin-friction coefficient compared against the reference data of Simeonides et al. \cite{simeonides1994experimental}.}
    \label{M6_ramp}
\end{figure}

Figure \ref{M6_ramp} (a) illustrates the density contours along with typical flow characteristics associated with the shock wave boundary layer interaction on a compression ramp. A weak oblique shock is generated at the leading edge of the plate as the laminar boundary layer grows spatially along the compression ramp. The results effectively capture the separation shock upstream of the ramp, the interaction between the shock and the boundary layer, and the reattachment of the flow downstream of the ramp. Figure \ref{M6_ramp} (b) shows the contours of the artificial bulk viscosity ($\beta^*$) with the LAD-C6 scheme. Higher values of $\beta^{*}$ are clearly evident across the shock discontinuities. Figures \ref{M6_ramp} (c) and (d) plot the streamwise evolution of the skin-friction coefficient ($C_f = 2\tau_w/(\rho_{\infty}u_{\infty}^2)$) and the wall-pressure coefficient ($K_p = (P-P_{\infty})/(0.5\rho_{\infty}u_{\infty}^2)$) for various schemes. The results of LAD-C6, MEG6 and MIG4 compare favourably with the reference data of Simeonides et al. \cite{simeonides1994experimental}.

\subsubsection*{2D Incident oblique shock-laminar boundary layer interaction}

The interaction of an incident oblique shock with a laminar boundary layer on an adiabatic flat plate is considered. The computational setup is in line with the numerical studies of Katzer \cite{katzer1989lengthscales}. The free-stream Mach number is $M_{\infty} = 2.0$ and the Reynolds number based on free-stream properties and the reference length scale, $x_{o}$, is $Re_{x_o} = 3 \times 10^5$. The oblique shock impinges at $x_{o}$ = 169.1$\delta^{*}$ in the inviscid case at an incident angle of $\beta=32.58^{\circ}$, where $\delta^{*}$ is the displacement thickness of the laminar boundary layer at $x_{o}$. The pressure ratio across the impinging shock and reflection is $p_{3}/p_{1} = 1.40$. The computational domain spanning $L_x \times L_y = 222\delta^{*} \times 100\delta^{*}$ is discretised using $N_{x} \times N_{y}$ = $300 \times 100$ in the streamwise and wall-normal directions respectively. An inflow laminar boundary layer profile is prescribed at 83.9$\delta^{*}$ upstream of the shock impingement position. A no-slip adiabatic wall condition is imposed on the bottom wall with variables extrapolated at the top and outlet boundaries.

\begin{figure}
    \centering
    \includegraphics{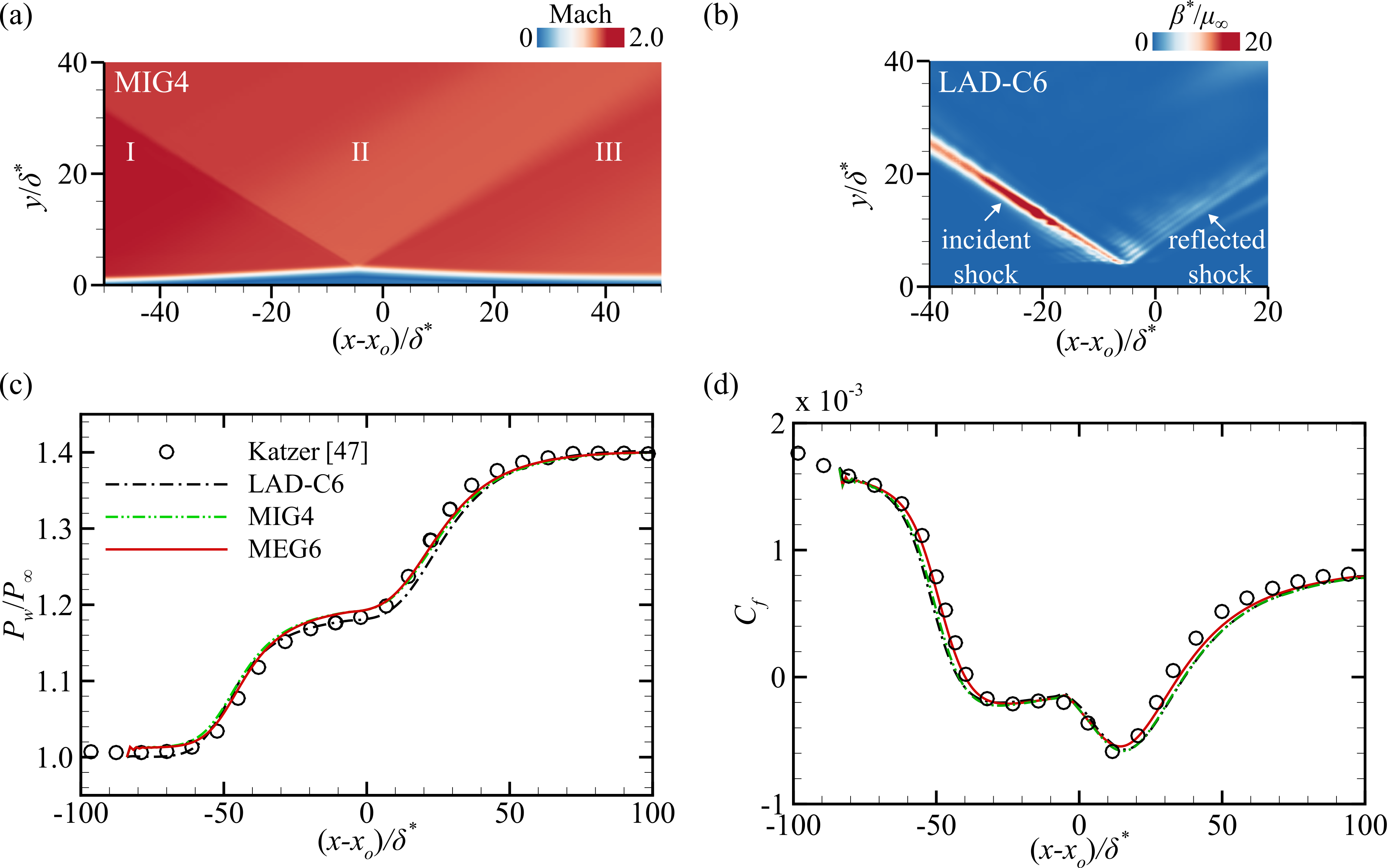}
    \caption{Contours of (a) Mach number with MIG4 scheme and (b)  artificial bulk viscosity ($\beta^*)$) across the shock with LAD-C6 scheme; streamwise evolution of (c) wall pressure and (d) skin-friction coefficient compared against the reference data of Katzer \cite{katzer1989lengthscales}.}
    \label{M2_obsbli}
\end{figure}

Figure \ref{M2_obsbli} (a) illustrates the contours of the Mach number. The shock structures, including incident oblique shock, reflected shock, and expansion waves, are accurately captured. Figure \ref{M2_obsbli} (b) demonstrates the distribution of the bulk viscosity  ($\beta^{*}$) across the shock discontinuities for the LAD-based simulation. Figures \ref{M2_obsbli} (c) and (d) demonstrate consistent predictions of the streamwise evolution of the wall pressure and skin friction coefficient across all schemes and favourable predictions against the reference numerical results of \textcolor{black}{Katzer} \cite{katzer1989lengthscales}.

For all 2D laminar test cases examined so far, the LAD and C-GBR schemes exhibited comparable accuracy. A comprehensive assessment in three-dimensional scenarios involving shocks and turbulence is essential and is presented in further detail in the next section.

\subsection{Three-dimensional viscous test cases}

In this section, we simulate 3D test cases that include a supersonic boundary layer on a flat plate and shock wave boundary layer interactions in supersonic and hypersonic flows. In all these simulations, it is crucial to impose realistic, coherent turbulent structures at the inlet. A digital filter-based (DF) turbulence generator \cite{veloudis2007novel, xie2008efficient, touber2009large} is specified at the inflow boundary. To account for the spatially varying turbulence scales, a multi-zone inflow technique is employed, where the inflow plane is divided into two zones in the wall-normal direction. Each velocity component ($u$, $v$, and $w$) is assigned a distinct integral length scale, based on values reported in the literature \cite{touber2009large, morgan2011improving}. Table \ref{df_scales} summarises the integral length scales used for all 3D simulations discussed in this section. Here, $I_x$, $I_y$ and $I_z$ are the integral length scales in the streamwise, wall-normal and spanwise directions, respectively. 

\begin{table}[width=0.6\linewidth,cols=4,pos=h]
\caption{Integral length scales for digital filtering}
\label{df_scales}
\begin{tabular*}{\tblwidth}{@{} LCCC @{}}
\toprule
Velocity component & $u$ & $v$ & $w$ \\
\midrule
$I_x / \delta_0$ & 0.7 & 0.28 & 0.28 \\
$I_y / \delta_0$ & 
    $0.13^{\text{inn}} - 0.23^{\text{out}}$ & 
    $0.17^{\text{inn}} - 0.3^{\text{out}}$ & 
    $0.1^{\text{inn}} - 0.13^{\text{out}}$ \\
$I_z / \delta_0$ & 0.15 & 0.15 & 0.3 \\
\bottomrule
\end{tabular*}

\vspace{1mm}
\textit{Note:} \textbf{inn} if $y \le y_{\rm lim}$, \textbf{out} if $y > y_{\rm lim}$, where $y_{\rm lim} = 1 \delta_1^{\rm vd}$
\end{table}

where $\delta_1^{\text{vd}}$ is the displacement thickness at the inflow boundary, computed from the van Driest–transformed velocity profile using the incompressible-flow definition. The digital filtering approach requires the specification of Reynolds stresses and mean flow properties such as velocity, density, and temperature. These quantities can be obtained from experimental measurements or high-fidelity simulations such as DNS or LES at comparable Reynolds numbers. Additionally, mean flow properties can also be obtained from preliminary RANS simulations or semi-empirical relations \cite{touber2009large} in the absence of high-fidelity data. In a compressible turbulent boundary layer, the fluctuations of thermodynamic variables (temperature and density) are non-trivial. \textcolor{black}{Touber and Sandham} \cite{touber2009large} recommended the use of the Strong Reynolds Analogy (SRA) to introduce thermodynamic fluctuations at the inlet. Although we were able to successfully test this strategy in supersonic flows, the use of SRA in hypersonic flows resulted in the divergence of the solution due to large fluctuations (both positive and negative) in temperature and density at the inlet. To mitigate the divergence issues that arise from this, two strategies can be considered: (a) avoid specifying thermodynamic fluctuations at the inlet and allow them to evolve physically within the computational domain or (b) impose the thermodynamic fluctuations using SRA estimated at a lower Mach number ($M\approx4$) that are almost realistic and allow them to develop physically in the computational domain. For the hypersonic test case, we have adopted the second strategy. For all three test cases, the simulation is performed for twelve flow-throughs, wherein one flow-through refers to the time taken by the fluid parcel to traverse the entire streamwise length of the domain. The first five flow-throughs are discarded to flush out transients, while the time averaging is performed for the subsequent ones. We have ensured that the time-averaged flow statistics, including Reynolds stresses, are statistically converged.

\subsubsection*{LES of Supersonic Turbulent Boundary Layer}

A large eddy simulation (LES) of a supersonic turbulent boundary layer is performed at a free-stream Mach number of $M_{\infty}$ = 1.3 and a Reynolds number based on the inlet momentum thickness and the free-stream viscosity of $Re_{\theta}$ = 1181. The computational setup aligns with the numerical studies by \textcolor{black}{Pirozzoli and Bernardini} \cite{pirozzoli2011turbulence}. The computational domain extends $L_x$ = 297$\theta_{in}$ in the streamwise direction, $L_{y}$ = 79.2$\theta_{\text{in}}$ in the wall-normal direction, and $L_z$ = 63.4$\theta_{\text{in}}$ in the spanwise direction, where $\theta_{\text{in}}$ is the momentum thickness of the inlet boundary layer. The flow is simulated on two different grids, comprising $N_x \times N_y \times N_z$ = $768 \times 256 \times 128$ and $512 \times 192 \times 96$, to demonstrate the effect of mesh resolution. The coarse mesh (G1) and the fine mesh (G2)  comprise 9.4 million and 25.2 million grid points, respectively. The reference DNS \cite{pirozzoli2011turbulence} employed a grid of 65.7 million points, which is nearly $2.6\times$ larger than the fine mesh used in the present study. The inner scale grid resolution is computed as $\Delta x^+ = u_{\tau}\Delta x / \nu_{w}$, where $u_{\tau}$ is the friction velocity and $\nu_{w}$ is the wall kinematic viscosity. The grid resolutions in the inner wall units for coarse mesh (G1) are $\Delta x^{+} = 19.1$,  $\Delta y^{+}_{w} = 0.92$,  $\Delta, z^{+} = 21.9$, while  $\Delta x^{+} = 12.7$, $\Delta y^{+}_{w} = 0.92$, and $\Delta z^{+} = 16.4$ for the fine mesh (G2). As discussed in the previous section, a digital filter-based inflow turbulence technique is specified at the inflow boundary. A no-slip, adiabatic wall condition is imposed on the bottom wall, while periodicity is enforced along the spanwise direction. The variables are extrapolated at the top and outlet boundaries. 

\begin{figure}
    \centering
    \includegraphics{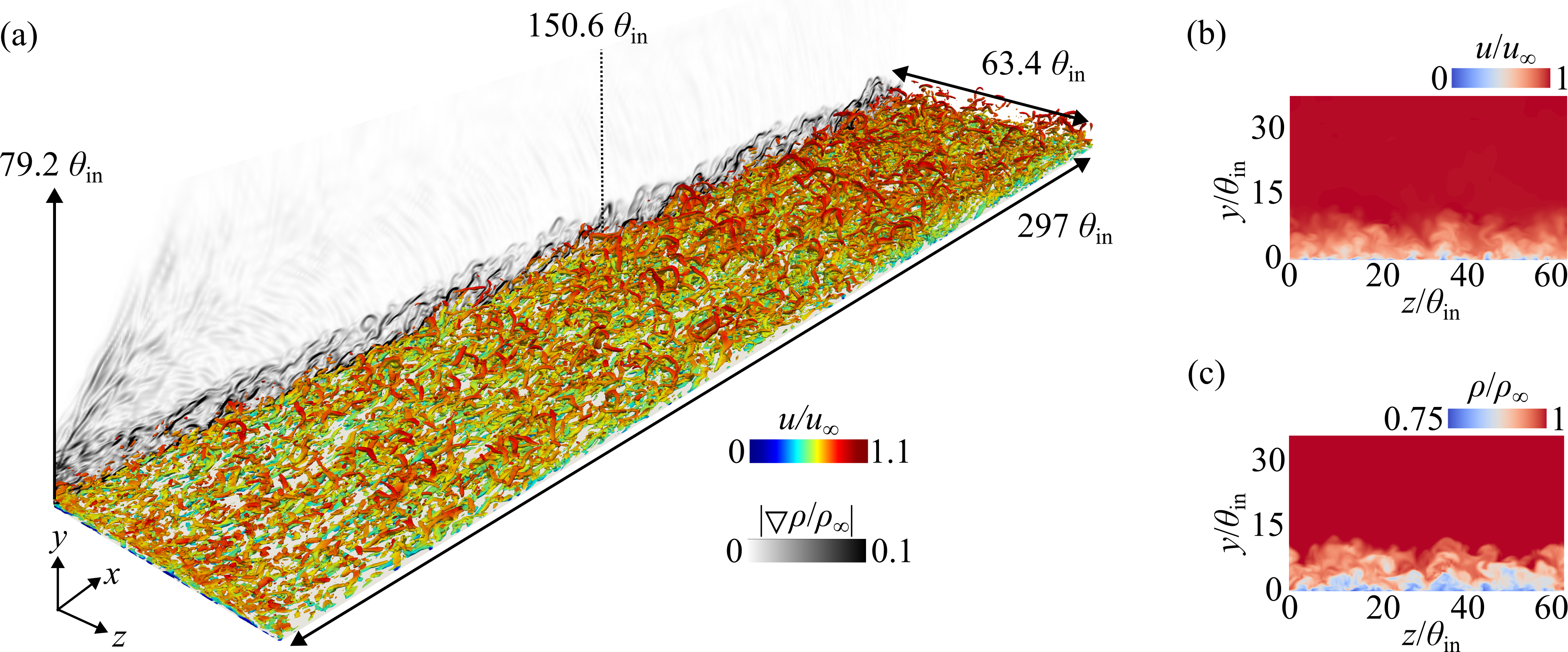}
    \caption{M1.3 - (a) Isosurfaces of Q-criterion coloured by streamwise velocity component with density gradient shown along the $xy$ plane. Instantaneous contours of the (b) streamwise velocity, and (c) density shown along the $yz$ plane at $x/\theta_{\text{in}}=150.6$.}
    \label{M1.3qiso}
\end{figure}

Figure \ref{M1.3qiso} (a) shows the isosurfaces of the Q criterion coloured with the streamwise component of the velocity. The isosurfaces illustrate the spatial growth of a non-equilibrium turbulent boundary layer to a fully developed state. The irregular vortical structures gradually transition to coherent structures, including hairpin vortex and fine-scale turbulent features. The magnitude of the density gradient is also visualised along the $xy$ plane in greyscale. A weak oblique shock is observed to emanate from the inlet as a result of the finite thickness of the inlet boundary layer. Moreover, weak compression waves are observed that decay along the streamwise direction. As noted by \textcolor{black}{ Adler et al.} \cite{adler2018synthetic}, this spurious noise is produced from the non-equilibrium boundary layer during the early relaxation length and occurs in both recycling/rescaling and synthetic turbulence generator techniques. Figures \ref{M1.3qiso} (b) and (c) illustrate the instantaneous contours of streamwise velocity and density at $x/\theta_{in}=150.6$, highlighting the three-dimensional nature of the flow.

In the literature, the LES data are largely compared with established incompressible correlations, as direct measurements of these features in the supersonic regime are scarce. The standard practice is to use the van Driest II transformation \cite{van1956problem} to compare the skin friction coefficients at different Mach numbers as follows:

\begin{equation}
    C_{fi} = F_{c}C_{f}, \quad Re_{\theta i} = \frac{\mu_\infty}{\overline{\mu}_w}Re_{\theta} = Re_{\delta_2} 
\end{equation}

where, the subscript $i$ denotes the incompressible values. The term $F_{c}$ for an adiabatic wall can be calculated as follows:
\begin{equation}
    F_{c} = \frac{\overline{T}_w/T_{\infty} - 1}{arcsin^2 \alpha}, \quad \alpha = \sqrt{1-\frac{T_{\infty}}{\overline{T}_w}}
\end{equation}

\begin{figure}
    \centering
    \includegraphics{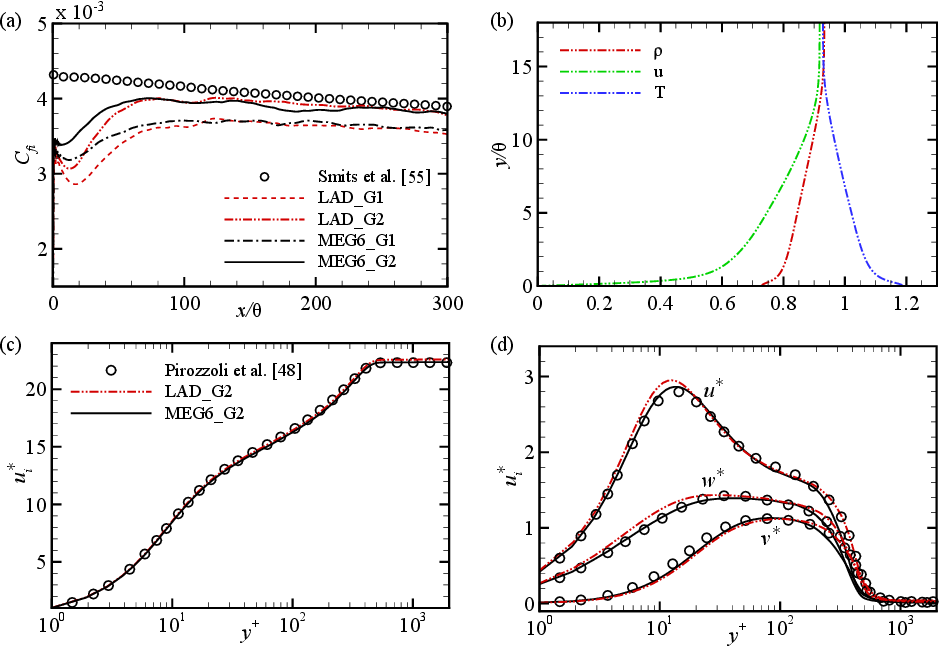}
    \caption{M1.3 - Profiles of (a) streamwise variation of van Driest II scaled skin friction coefficient, (b) streamwise velocity, density, and temperature at $x/\theta_{\text{in}}=150.6$, (c) Van Driest transformed mean velocity in wall units, and (d) density scaled Reynolds stress components extracted at $Re_{\delta_{2}}$ = 1327.}
    \label{M1.3plot}
\end{figure}

Figure \ref{M1.3plot} compares the van Driest II (VDII) transformed skin friction coefficient \cite{van1956problem} with the theoretical correlations of Smits et al. \cite{smits1983low}. It is evident that $C_{f_{i}}$ deviates from the VDII prediction in the initial recovery section, from the inlet up to $x = 120\theta_{\text{in}}$. This discrepancy arises because the inlet boundary layer requires additional development length before evolving into a fully developed turbulent state. The coarse mesh (G1) results are consistently under-predicted, while the results on fine mesh (G2) compare favourably with the VDII prediction across both the LAD-C6 and MEG6 schemes. Figure \ref{M1.3plot} (b) plots the boundary layer profiles of streamwise velocity $u$, density $\rho$, and temperature $T$ extracted at $x=150.6\theta_{\text{in}}$, which can be used to set up the simulation using appropriate boundary-layer scaling. Figure \ref{M1.3plot} (c) compares the van Driest transformed mean velocity ($u^+_\text{VD}$) with the fine mesh G2. All profiles were extracted at $Re_{\delta_{2}}$ = 1327, where $Re_{\delta_{2}}$ is the Reynolds number based on momentum thickness and wall viscosity. The LAD-C6 and MEG6 profiles are in excellent agreement with the reference DNS results \cite{pirozzoli2011turbulence} in all regions of the velocity profile. Figure \ref{M1.3plot} (d) compares the density-scaled Reynolds stresses ($u^*_i = \sqrt{\overline{\rho}/{\overline{\rho}_w}}  \ \overline{u_i'}/{u_{\tau}}$) with the fine mesh G2. Despite a $2.6\times$ coarser grid than DNS, the results of the LAD-C6 and MEG6 schemes agree well with the reference DNS data. However, marginal overprediction of Reynolds stresses in streamwise ($u^*$) and spanwise ($w^*$) is observable in the buffer layer with the LAD-C6 scheme.

\subsubsection*{LES of M2.9 Compression Ramp}

\begin{figure}
    \centering
    \includegraphics{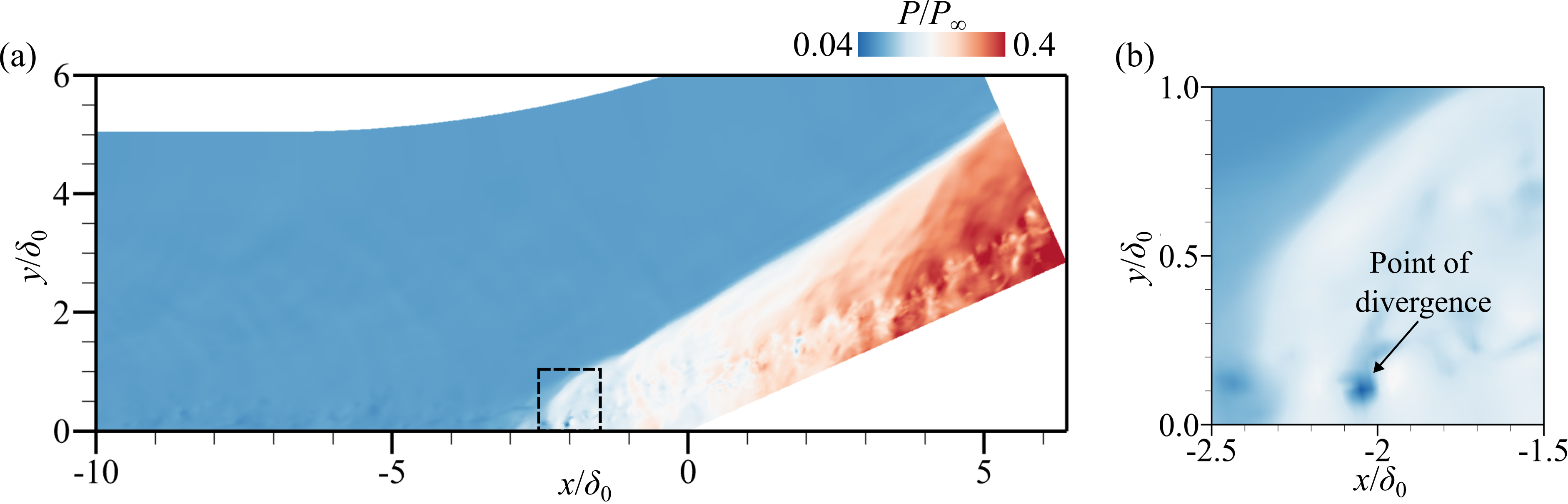}
    \caption{M2.9 - Contours of density using LAD-C6 scheme (a) illustrating the flow-field and (b) a zoomed-in view showing the problematic cells prior to the onset of abrupt divergence.}
    \label{M2.9_lad}
\end{figure}

A large eddy simulation (LES) of a supersonic compression ramp is carried out at a free-stream Mach number of $M_{\infty}$ = 2.9 and a compression corner angle of $\theta_c = 24^{\circ}$. The Reynolds number based on the inlet momentum thickness and the free-stream viscosity is $Re_{\theta}$ = 1800. The computational setup aligns with the numerical studies conducted by Kokkinakis et al. \cite{kokkinakis2020direct}. The extent of the computational domain is $L_x$ = 534.6$\theta_{\text{in}}$, $L_{y}$ = 161$\theta_{\text{in}}$ and  $L_z$ = 61.2$\theta_{\text{in}}$ in the streamwise, wall-normal, and spanwise direction, respectively. While $\delta_0$ is used as the reference length scale in Kokkinakis et al. \cite{kokkinakis2020direct}, we adopt $\theta_\text{in}$ as the reference in the present study, with $\theta_\text{in}/\delta_0 \approx 20.4.$ Simulation is carried out on grid size $N_x \times N_y \times N_z$ = $523 \times 167 \times 120$ and $681 \times 167 \times 151$, comprising 17.2 million grid points. The grid resolutions in the inner wall units is $\Delta x^{+} = (6.4 - 14.2 )$, $\Delta y^{+}_{w} = 0.85$, and $\Delta z^{+} = 5.8$. A digital filter-based inflow turbulence technique is specified at the inflow boundary. A no-slip, isothermal wall condition, $T_{w}/T_{r} = 1.14$, is imposed on the bottom wall, where $T_{r}$ is the recovery adiabatic wall temperature. A supersonic outflow boundary condition is maintained at the outlet, while the flow variables are extrapolated at the top wall. Periodicity is imposed in the spanwise direction. 

\begin{figure}
    \centering
    \includegraphics{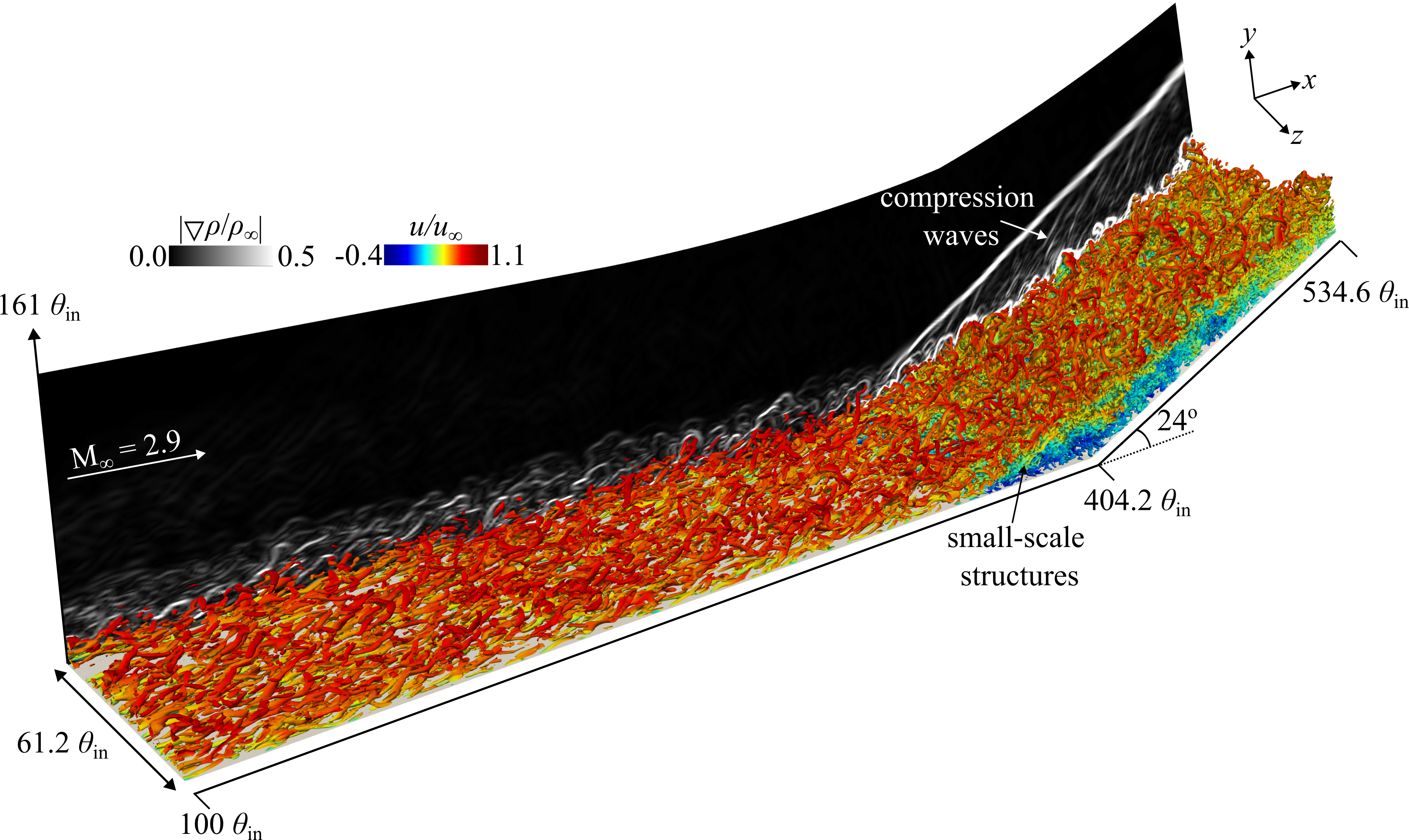}
    \caption{M2.9 - Isosurfaces of Q-criterion coloured by streamwise velocity component with density gradient shown along the $xy$ plane}
    \label{M2.9_qiso}
\end{figure}

Upon simulating the test case with the LAD-C6 scheme, it eventually encountered negative density and pressure values at a few grid points, resulting in abrupt divergence. Figure \ref{M2.9_lad} illustrates the density contours along with a magnified region showing the problematic cells before the onset of abrupt divergence. Various measures were attempted to stabilise the simulation, including a decrease in $\alpha_f$ and the filter order, a decrease in the time step and an increase in the dimensionless artificial bulk viscosity coefficient ($C_\beta$), but the schemes did not produce a stable simulation. Hence, the results are shown only with the MEG6 scheme. We have further re-attempted this case using a hybrid solver that will be discussed in Section \ref{hybrid}. Figure \ref{M2.9_qiso} shows the instantaneous isosurfaces of Q-criterion coloured by the streamwise velocity component spanning across $x = 100\theta_{\text{in}}$ to the outlet plane. The coherent streamwise structures are dominant in the flat plate region and intensify as the flow separates due to the adverse pressure gradient near the compression ramp. These streamwise vortices break down into smaller ones in the separation region beyond the strong oblique shock across the compression ramp. The contours of the density gradient are also shown on a 2D $xy$ plane, highlighting regions of significant density variation. In addition, compression waves are clearly visible within the shock and boundary layer regions. 

\begin{figure}
    \centering
    \includegraphics{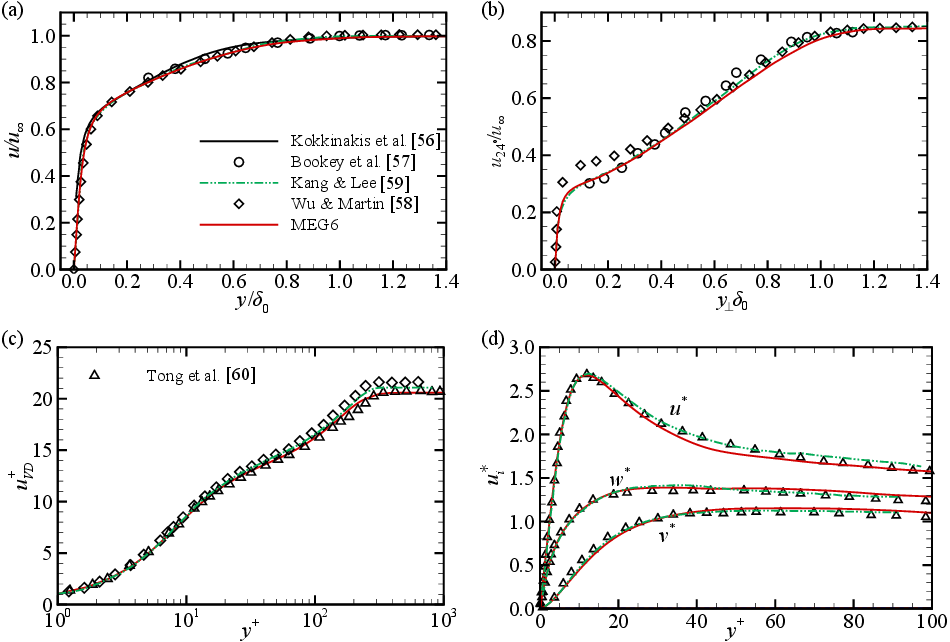}
    \caption{M2.9 - Profiles of (a) mean streamwise velocity at $x = -8\delta_\text{0}$, (b) mean streamwise velocity at $x = 4\delta_\text{0}$, (c) Van Driest transformed mean velocity in wall units, and (d) density scaled Reynolds stress components extracted at $x = -5.4\delta_\text{0}$.}
    \label{M2.9plot1}
\end{figure}

Figure \ref{M2.9plot1} (a) plots the mean streamwise velocity profile extracted at $x = -8\delta_\text{0}$, where the Reynolds number based on the momentum thickness and the free-stream viscosity is  $Re_{\theta}$ = 2300. The present results show excellent agreement with the experimental data of Bookey et al. \cite{bookey2005new} and various other numerical studies \cite{kokkinakis2020direct,  wu2007direct, kang2024direct}. Figure \ref{M2.9plot1} (b) plots the mean tangential velocity at $4\delta_\text{0}$ downstream of the compression corner, where $u_{24^\circ} = (u \cos(24^\circ) + v \sin(24^\circ))$ and $y_{\perp}$ is the wall-normal distance. The plot clearly shows that the velocity profile computed with the MEG6 scheme matches closely with both experimental and numerical studies \cite{ bookey2005new, wu2007direct, kang2024direct}. Figures \ref{M2.9plot1} (c) and (d) further plot the wall-normal variation of mean Van Driest transformed velocity ($u^{+}_{VD}$) and density-scaled Reynolds stress components ($u_i^*$) in viscous units against $y^{+}$. The LES predictions are also compared with the DNS results reported in the literature \cite{wu2007direct, kang2024direct, tong2017numerical}. $u^{+}_{VD}$ closely align with the DNS results in the viscous sublayer and the logarithmic region. Moreover, the present LES shows good agreement in $u_i^*$ with previous studies. The streamwise velocity fluctuation ($u^*$) exhibits a pronounced peak at $y^{+} \approx 12$, consistent with DNS data.

\begin{figure}
    \centering
    \includegraphics{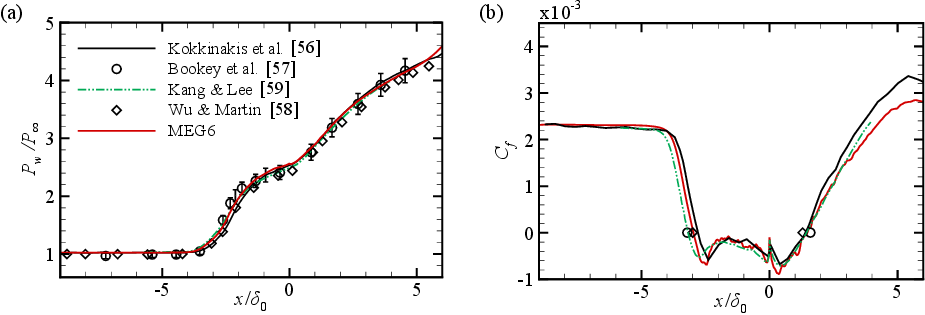}
    \caption{M2.9 - Profiles of streamwise variation of (a) wall pressure, and  (b) skin friction coefficient computed using MEG6 scheme and compared against the reference data.}
    \label{M2.9plot2}
\end{figure}

Figures \ref{M2.9plot2} (a) and (b) compare the streamwise evolution of wall pressure ($P_w$) and skin friction coefficient ($C_f$) with DNS results \cite{ wu2007direct, kokkinakis2020direct, kang2024direct} and the experimental data of Bookey et al. \cite{bookey2005new}. The predictions of $P_w$ with the MEG6 scheme agree favourably with the DNS results and are within the experimental error bounds reported in the literature. In the $C_f$ plot, the separation and reattachment points from the data of Bookey et al. \cite{bookey2005new} and Wu \& Martin \cite{wu2007direct} are also indicated. A sharp decrease in $C_{f}$ is observed upstream of the separation point, reaching negative values within the separated region, followed by a recovery downstream of reattachment. It is encouraging to note that the present $C_f$ results and the extent of separation across the compression corner agree well with the experimental and DNS data.

\subsubsection*{LES of M7.2 Compression Ramp}

\begin{figure}
    \centering
    \includegraphics{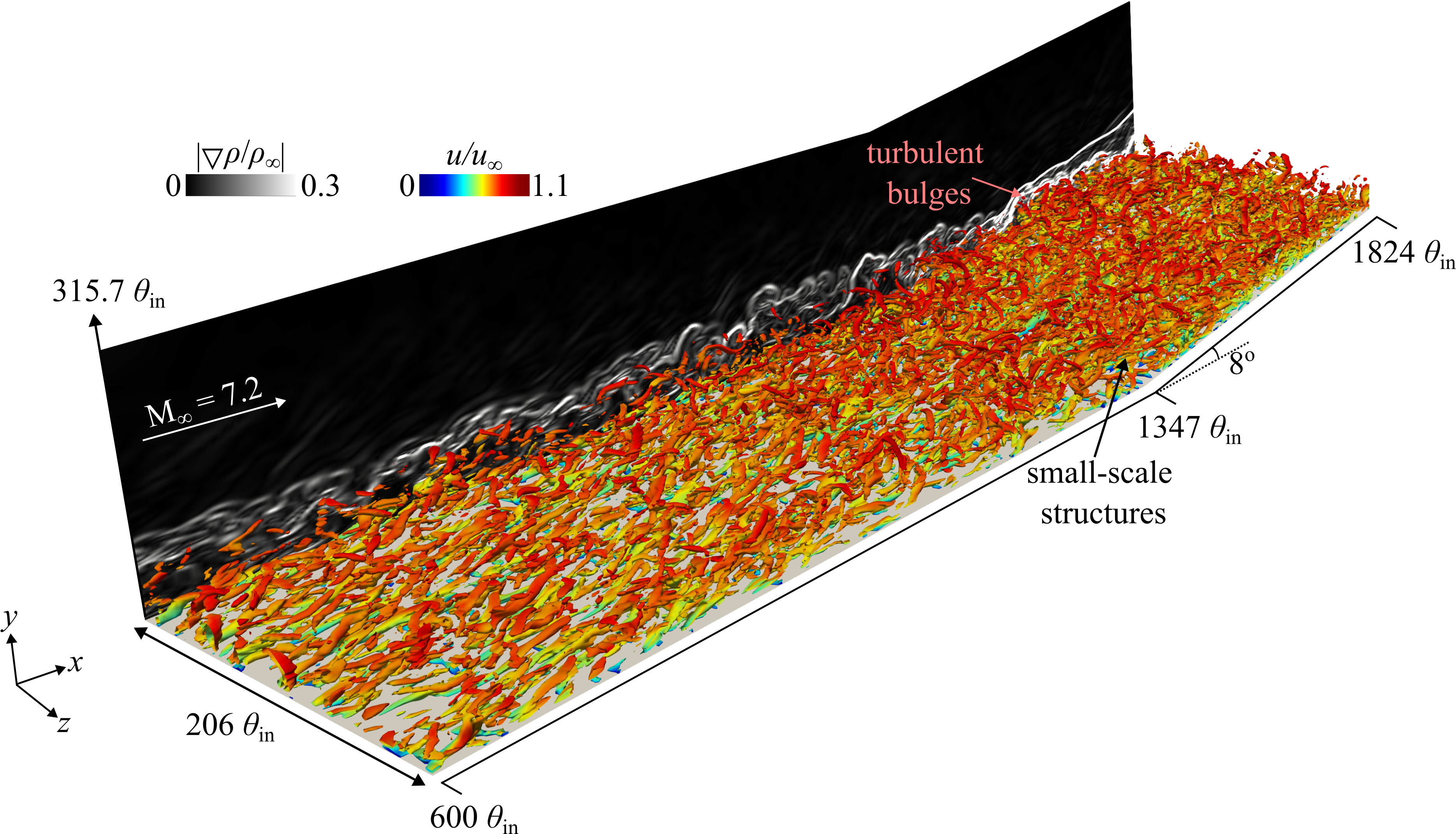}
    \caption{M7.2 - Isosurfaces of Q-criterion coloured by streamwise velocity component with density gradient shown along the $xy$ plane.}
    \label{M7_qiso}
\end{figure}

A large eddy simulation (LES) of a hypersonic compression ramp is performed at a free-stream Mach number of $M_{\infty}$ = 7.2 and compression corner angle of $\theta_c = 8^{\circ}$. The Reynolds number based on the inlet momentum thickness ($\theta_{\text{in}}$) and the free-stream viscosity is $Re_{\theta}$ = 3300. The computational setup is in line with the numerical studies by Priebe \& Martin \cite{priebe2021turbulence}. Since the ambient temperature $T_{\infty}=62.9$ K, the authors have used Keyes' law instead of Sutherland's for calculating dynamic viscosity, as the latter resulted in an error of $\approx7.5\%$ at such low temperatures. The Keyes' law to calculate dynamic viscosity, is $\mu = 1.488 \times 10^{-6}\sqrt{T}/{(1 + \left(122.1/T\right) \times 10^{-5/T})}$, where $\mu$ is in Pa s and $T$ is the temperature in Kelvin. The extent of the computational domain is  $L_x$ = 1824$\theta_{\text{in}}$, $L_{y}$ = 315.7$\theta_{\text{in}}$ and $L_z$ = 206$\theta_{\text{in}}$  in the streamwise, wall-normal, and spanwise direction.  While $\delta_\text{r}$ is used as the reference length scale in Priebe \& Martin \cite{priebe2021turbulence}, we adopt $\theta_\text{in}$ as the reference in the present study, with $\theta_\text{in}/\delta_r \approx 45.5$. Simulation is carried out on a grid size $N_x \times N_y \times N_z$ = $738 \times 171 \times 121$, comprising 15.2 million grid points. The grid resolutions in the inner wall units is $\Delta x^{+} = (8.4 - 16.9)$, $\Delta y^{+}_{w} = 0.9$, and $\Delta z^{+} = 10.6$. It is worth noting that the reference paper has employed a recycling-rescaling based technique while the present study exploits a digital filter approach that prescribes the inlet turbulent boundary layer. A no-slip, isothermal wall condition, $T_{w}/T_{r} = 5.40$, is imposed on the bottom wall, where $T_{r}$ is the recovery adiabatic wall temperature. Periodicity is imposed in the spanwise direction, while the flow variables are extrapolated at the top and outlet boundaries. 

\begin{figure}
    \centering
    \includegraphics{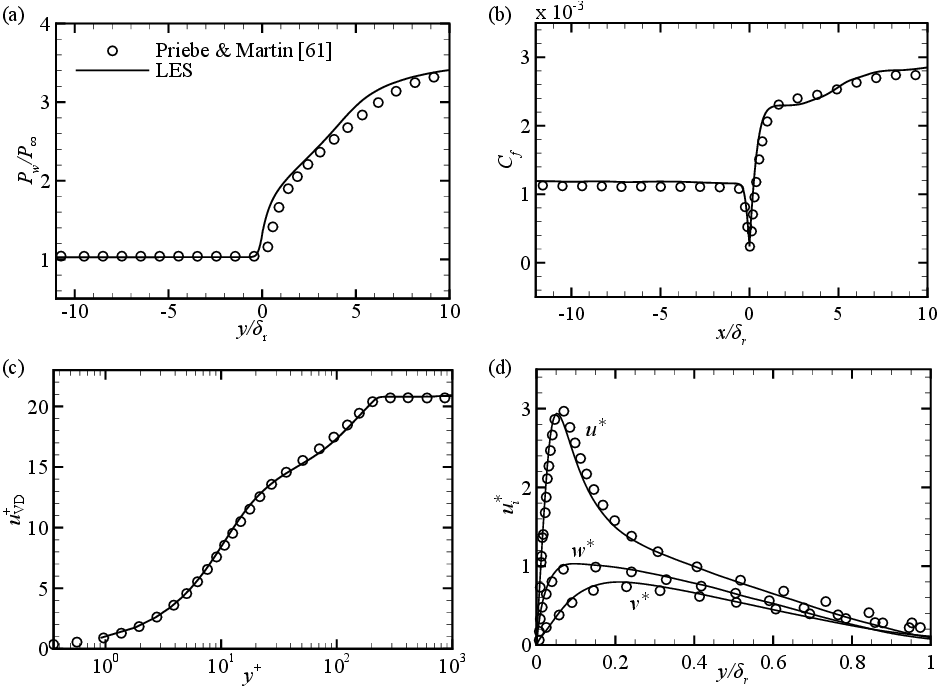}
    \caption{M7.2 - Plots of (a) Streamwise evolution of Wall pressure, (b) Streamwise evolution of skin friction coefficient, (c) Van Driest transformed mean velocity, and (d) Reynolds stress components in wall units from LES results compared with the reference DNS data \cite{priebe2021turbulence}.}
    \label{M7_stats}
\end{figure}

The simulation with the LAD-C6 scheme produced unphysical negative density and pressure values at a few grid points and diverged abruptly. Hence, only the results from the MEG6 scheme are presented. Figure \ref{M7_qiso} shows the instantaneous isosurfaces of Q-criterion coloured by the streamwise velocity component spanning across $x/\theta_{\text{in}} = 600$ to the exit plane. Similar to the previous supersonic case, the isosurfaces show increased turbulence levels at the compression corner when the shock interacts with the boundary layer. The density gradient is also shown along the $xy$ plane, indicating the location and thickness of the shock. A slight distortion in the shock structure is also apparent, formed as a result of turbulent bulges from the upstream boundary layer. Figures \ref{M7_stats} (a) and (b) demonstrate the streamwise variation of wall pressure ($P_w$) and skin friction coefficient ($C_f$) predicted using the MEG6 scheme. It is evident that the predictions agree well with the reference DNS data of Priebe \& Martin \cite{priebe2021turbulence}. The pressure rises marginally upstream of the compression corner and continues to increase through the interaction region with gradual recovery further downstream. The value of $P_w/P_{\infty}$ at the outlet plane is \textit{3.45}, which is close to the inviscid shock limit of 3.51. In the $C_f$ plot, the skin friction coefficient decreases significantly upstream of the corner but remains positive, indicating the absence of flow separation. Figures \ref{M7_stats} (c) and (d) further compare the van-Driest transformed mean velocity ($u_{\text{VD}}^+$) and Reynolds stresses $(({\overline{u'^2}})^{0.5}/u_\tau)$. The wall-normal profiles are extracted at $x = 800\theta_{in}$, which corresponds to the inflow plane of the reference simulation \cite{priebe2021turbulence}. The profile of $u_{VD}^+$ agrees well with the reference data in both the log-layer and the wake region. The Reynolds stress trends are well captured and align with the DNS results, thus demonstrating the robustness of the C-GBR framework.

\section{Test Cases Overview and Speed-up Analysis} \label{overview}

\subsection{Summary of Test Cases} \label{summarize}

Table \ref{tbl:scheme_comparison} summarises the test cases simulated using LAD-C6  and centralised gradient-based reconstruction schemes (C-GRB: MEG6/MIG4). A diverged simulation is represented by an x-mark (\xmark) while a check-mark (\cmark) indicates a converged run. The comparative analysis shows contrasting differences between the two numerical approaches. Both schemes performed equally well with smooth or moderately challenging problems, including the Shu-Osher problem, double periodic shear layer, 2D laminar hypersonic ramp, 2D oblique shock boundary layer interaction and LES of M1.3 turbulent boundary layer. While LAD-C6 performs satisfactorily with the majority of the benchmark cases, it fails to provide a stable simulation with cases including strong discontinuities. Specifically, the LAD-C6 schemes fail with test cases such as the Le-Blanc problem, strong 2D Riemann problem, compressible triple points and 3D problems, including large eddy simulation (LES) of M2.9 and M7.2 compression ramp. In stable cases, LAD-C6 marginal oscillations are observed while resolving contact discontinuity in the Sod shock tube and Lax problems. Moreover, the schemes predict relatively thicker shocks, which is evident in the Richtmeyer Meshkov instability and shock-bubble interaction cases. In contrast, centralised gradient-based reconstruction schemes successfully simulated all test cases, including those that failed with LAD-C6 schemes. The C-GRB schemes produced a stable and accurate solution with minimal numerical dissipation.
\begin{table}[width=0.95\linewidth,cols=5,pos=h]
\caption{Summary of the test cases simulated using LAD-C6  and centralised gradient based reconstruction schemes (C-GRB: MEG6/MIG4). An unsuccessful simulation is represented with an x-mark (\xmark) while a check-mark (\cmark) indicates a successful run.  CD stands for contact discontinuity.}
\label{tbl:scheme_comparison}
\begin{tabular*}{\tblwidth}{@{} l l c c p{5cm} @{}}
\toprule
\makecell{Test Case} & Cells & LAD-C6 & C-GRB & Remarks \\
\midrule
Sod Shock Tube                      & $200 \times 1$                  & \cmark & \cmark & LAD-C6: oscillations with CD \\
Lax Problem                         & $200 \times 1$                  & \cmark & \cmark & LAD-C6: oscillations with CD \\
Le-Blanc Problem                    & $900 \times 1$                  & \xmark & \cmark & LAD-C6: divergence  \\
Shu–Osher                           & $300 \times 1$                  & \cmark & \cmark & Similar accuracy with both schemes \\
Shock–Entropy Wave Test            & $400 \times 80$                 & \cmark & \cmark & LAD-C6: Similar accuracy with both schemes \\
Richtmyer–Meshkov Instability      & $320 \times 80$                 & \cmark & \cmark & LAD-C6: higher dissipation \\
Shock–Vortex Interaction           & $1024 \times 512$               & \cmark & \cmark & LAD-C6: Similar accuracy with both schemes \\
Kelvin–Helmholtz Instability       & $512 \times 512$                & \cmark & \cmark & LAD-C6: oscillations with CD \\
2D Riemann Problem (C3)            & $400 \times 400$                & \xmark & \cmark & LAD-C6: divergence \\
2D Riemann Problem (C12)           & $800 \times 800$                & \cmark & \cmark & LAD-C6: marginal oscillations \\
Compressible Triple Point          & $1792 \times 768$               & \xmark & \cmark & LAD-C6: divergence \\
Shock–Bubble Interaction           & $1300 \times 356$               & \cmark & \cmark & LAD-C6: higher dissipation \\
Double Periodic Shear Layer        & $320 \times 320$                & \cmark & \cmark & Similar accuracy with both schemes \\
2D Hypersonic Compression Ramp     & $378 \times 81$                 & \cmark & \cmark & Similar accuracy with both schemes \\
2D Oblique Shock Boundary Layer    & $300 \times 100$                & \cmark & \cmark & Similar accuracy with both schemes \\
LES of M1.3 TBL                   & \makecell{$512 \times 192 \times 96$\\$768 \times 256 \times 128$} & \cmark & \cmark & Similar accuracy with both schemes \\
LES of M2.9 Compression Ramp       & $681 \times 167 \times 151$    & \xmark & \cmark & LAD-C6: divergence \\
LES of M7.2 Compression Ramp       & $738 \times 171 \times 121$    & \xmark & \cmark & LAD-C6: divergence \\
\bottomrule
\end{tabular*}
\end{table}

\subsection{ Speed-up Comparison of Computational Methods}
In this section, the pseudocode of the numerical frameworks for LAD and C-GRB is described, along with the contribution of each subroutine to the total computational time. Subsequently, their computational efficiency is demonstrated. The pseudocode of the LAD and C-GRB framework is outlined in Algorithm \ref{LAD_algo} and \ref{GBR_algo}, respectively. With the LAD approach, the inviscid fluxes are first computed. The viscous routines involve calculating the gradients of primitive variables, followed by LAD routines that compute artificial transport coefficients and viscous fluxes. After the residual calculation (RESIDUAL\_CALC), the solution is advanced in time using NETFLUX\_LOOP. An implicit filtering operation (FILTER\_CALLS) is employed in the final Runge-Kutta stages to reduce the dispersive errors. The boundary conditions are then imposed along with a few auxiliary subroutines. In contrast, the C-GRB approach first computes the gradients of primitive variables. In the subsequent step, the primitive variables are reconstructed through limiting (PRIM\_HALF\_CELL), and the inviscid fluxes are calculated using a Riemann solver (RIEMANN\_FLUXES). The viscous fluxes are then computed using VISCOUS\_ROUTINES. All subsequent subroutines are identical to LAD, except for the filtering routines, as no filtering is required in C-GBR schemes.

\begin{algorithm} 
\caption{Main steps of the COMPSQUARE solver algorithm with LAD-based scheme}
\begin{algorithmic}[1]
\State Allocate variables, compute grid metric, initialisation and numerics 
\For{iter = 1 to $N_{\text{steps}}$} \Comment{Time iterations}
    \For{step = 1 to $N_{\text{rk}}$} \Comment{Runge-Kutta stages}
        \State \textsc{Inviscid\_routines} (Inviscid fluxes calculation)
        \State \textsc{Viscous\_routines} (Viscous fluxes calculation)
        \State \hspace{2em}\textsc{DiscretizationIJK} (Derivatives of primitive variables $U_i$, $U_j$, $U_k$, $\ldots$)
         \State \hspace{2em}\textsc{LAD routines} (Computing $\mu^*$, $\beta^*$ and $\kappa^*$)  
        \State \textsc{Residual\_calc} (computing flux derivatives)          
        \State \textsc{Netflux\_loop} (Advancing solution in time)
        \If{$i_{\text{rk}} = N_{\text{rk}}$}
        \State \textsc{Filter\_Calls} (Apply filtering along $i,\ j,\ k$)
        \EndIf
        \State \textsc{Set\_primitives} (Update primitive variables)
        \State \textsc{Set\_BCs} (Impose Boundary Conditions)
    \EndFor
\EndFor
\State \textsc{Output} (Save required statistics to file)
\end{algorithmic}  
\label{LAD_algo}
\end{algorithm}

\begin{algorithm}
\caption{Main steps of the COMPSQUARE solver algorithm with C-GRB schemes}
\begin{algorithmic}[1]
\State Allocate variables, compute grid metric, initialisation and numerics 
\For{iter = 1 to $N_{\text{steps}}$} \Comment{Time iterations}
    \For{step = 1 to $N_{\text{rk}}$} \Comment{Runge-Kutta stages}
        \State \textsc{Inviscid\_routines} (Inviscid fluxes calculation)
        \State \hspace{2em}\textsc{DiscretizationIJK} (Derivatives of primitive variables $U_i$, $U_j$, $U_k$, $\ldots$)
        \State \hspace{2em} \textsc{Prim\_Half\_Cell} (Reconstruction of primitive variable and limiting)
        \State \hspace{2em} \textsc{Riemann\_Fluxes} (Compute Fflux, Gflux, and Hflux using HLLC solver)
        \State \textsc{Viscous\_routines} (Viscous fluxes calculation)
        \State \textsc{Residual\_calc} (computing flux derivatives) 
        \State \textsc{Netflux\_loop} (Advancing solution in time)
        \State \textsc{Set\_primitives} (Update primitive variables)
        \State \textsc{Set\_BCs} (Impose Boundary Conditions)
    \EndFor
\EndFor
\State \textsc{Output} (Save required statistics to file)
\end{algorithmic}
\label{GBR_algo}
\end{algorithm}

\begin{figure}
    \centering
    \includegraphics{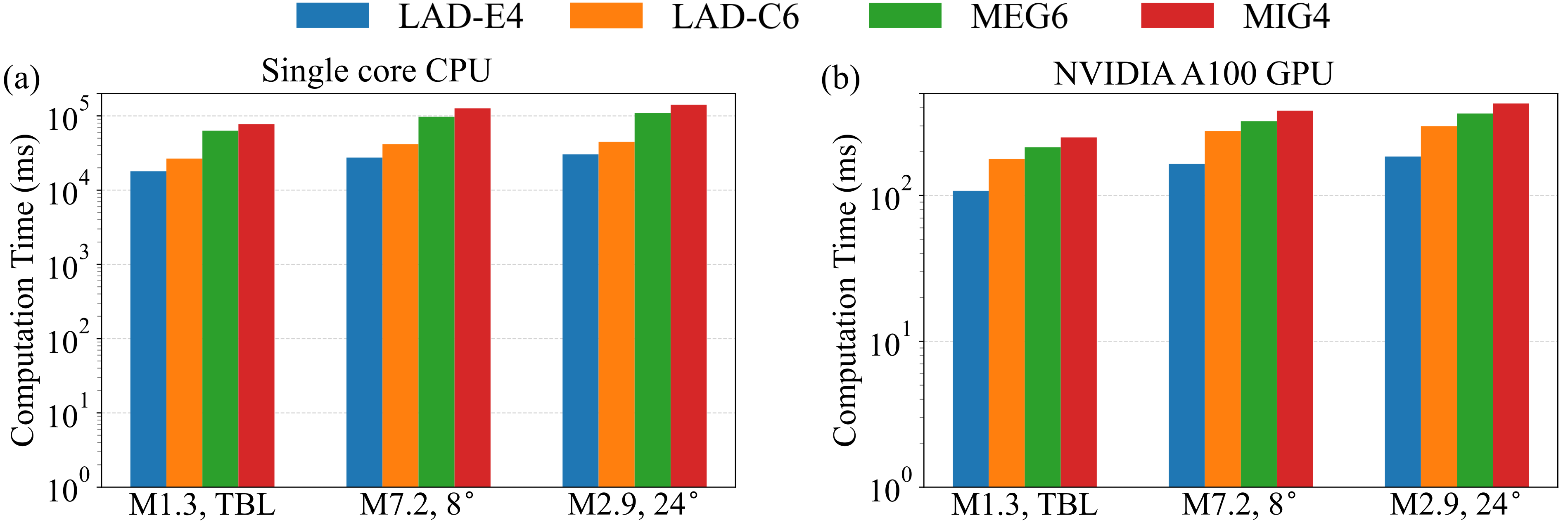}
    \caption{Computational time per RK stage (in ms) across different test cases with LAD-E4, LAD-C6, MEG6 and MIG4 schemes on (a) single-core CPU, and (b) NVIDIA A100 GPU.}
    \label{speedup}
\end{figure}

\begin{figure}
    \centering
    \includegraphics{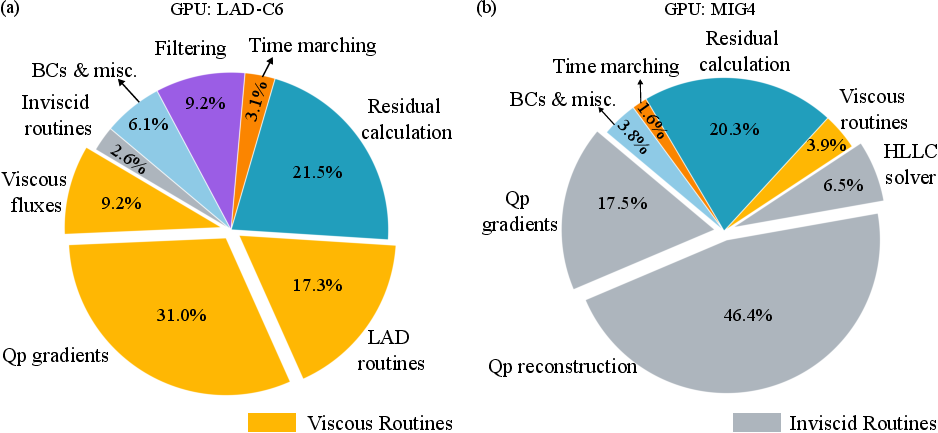}
    \caption{Breakdown of total computational time spent in various subroutines per RK stage executed on an NVIDIA A100 GPU using (a) LAD-C6 and (b) MIG4 scheme for the M7.2 compression ramp.}
    \label{sub_comp}
\end{figure}

The computational efficiency of these schemes is demonstrated on (a) M1.3 turbulent boundary layer comprising 9.4 million grid points, (b) M7.2 $8^{\circ}$ compression ramp comprising 15.3 million grid points and (c) M2.9 $24^{\circ}$ compression ramp comprising 17.2 million grid points. These test cases are selected to evaluate the performance of the numerical framework in various flow conditions and features. The computations are carried out in double precision (FP64) on a single-core Intel Xeon Gold 6240R and an NVIDIA A100 GPU with 80 GB HBM2e memory. In the present study, GPU parallelism is implemented using a directive-based OpenACC framework. Here, we report simulations on a single GPU, although the code is capable of scaling across multiple GPUs using both MPI and OpenACC. Moreover, the computational time of the LAD-E4 scheme is estimated to provide a comprehensive assessment, although its flow solution results are omitted and not discussed so far. Figures \ref{speedup} (a) and (b) compare the computational time per RK stage (in ms) across different test cases with LAD-E4 (explicit fourth-order), LAD-C6, MEG6 and MIG4 schemes on CPU and GPU. As expected, the results show that computations on the A100 GPU are approximately two orders of magnitude faster than those on the single-core CPU. For the M2.9 $24^{\circ}$ test case, the LAD-C6 and MEG6 schemes achieve speed-ups of $150\times$ and $301\times$, respectively, compared to single-core CPU computations. It is apparent from Figure \ref{speedup}(b) that the implicit schemes (LAD-C6 and MIG4) are computationally more expensive than the explicit schemes (LAD-E4 and MIG6) as the former involve solving several 1D tridiagonal matrices. In addition, LAD-based schemes are computationally more economical than C-GBR schemes. Specifically, LAD-C6 demonstrates a speed-up of $1.17-1.22\times$ and $1.38-1.43\times$ compared to the MEG6 and MIG4 schemes, while LAD-E4 is $1.9-2.0\times$ faster than MEG6 and $2.2-2.3\times$ faster than MIG4. However, both MEG6 and MIG4 schemes show significantly higher GPU acceleration compared to LAD schemes, achieving $2\times$ speed-ups relative to their respective CPU runtimes, thereby demonstrating their computational efficiency when leveraged on GPU architectures.


Figures \ref{sub_comp} (a) and (b) further illustrate the subroutine-level breakdown of the total computational time per RK stage executed on an NVIDIA A100 GPU using the LAD-C6 and MIG4 schemes for the M7.2 compression ramp test case. Both frameworks utilise the same runtime for the residual calculation, at 21.5\% in LAD-C6 and 20.3\% in MIG4. The time marching and boundary conditions, along with miscellaneous routines, contribute 9.2\% to the LAD-C6, while 5.4\% to the MIG4. In the case of LAD-C6, a significant contribution to runtime comes from viscous routines (57.5\%), which include the computation of the primitive variable (Qp) gradient, LAD routines, and the viscous fluxes. In contrast, MIG4 spends only 3.9\% of the total time on viscous routines. In contrast, the MIG4 scheme spends 70.4\% of the total runtime with inviscid routines, including Qp gradients, Qp reconstruction, and Riemann (HLLC) solver, while the inviscid fluxes account for 2.6\% with the LAD-C6 scheme. A notable difference between the two schemes is the presence of an implicit filtering routine in the LAD-C6 approach, which accounts for 9.2\% of the computational time. Although the filtering routine is absent in the MIG4 approach, the LAD-C6 scheme is still faster than the former.

\section{Coupled LAD$-$GBR Approach}  \label{hybrid}

\begin{figure}
    \centering
    \includegraphics{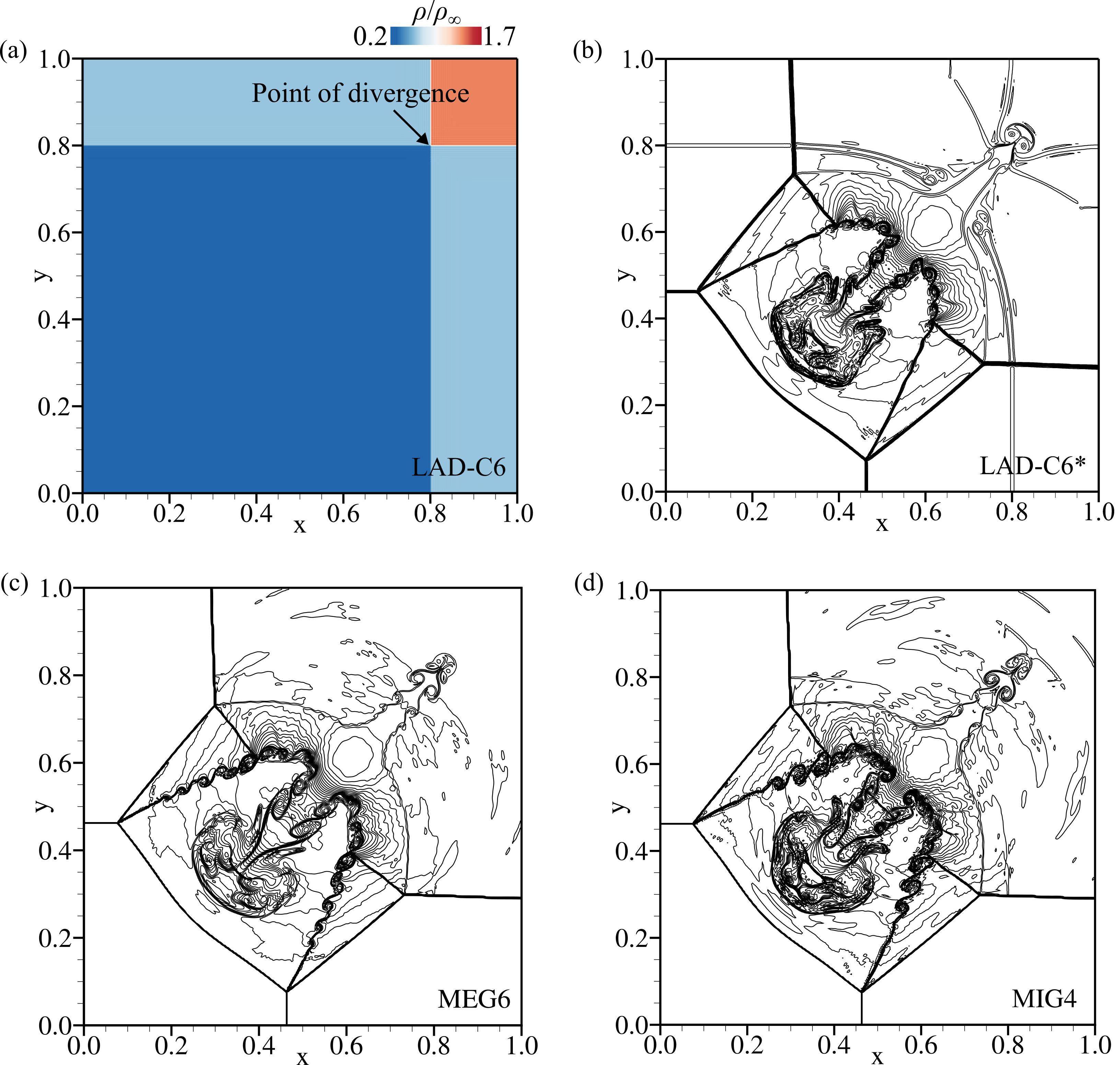}
    \caption{Density contour of the strong 2D Riemann problem with (a) LAD-C6 scheme showing problematic cells prior to the onset of the divergence; and 30 uniformly spaced contour levels with (b) LAD-C6$^*$, (c) MEG6, and (d) MIG4 scheme.}
    \label{riemann_hyb}
\end{figure}

As discussed in subsection \ref{summarize}, the LAD approach failed to converge for certain test cases. For example, the strong 2D Riemann test case diverged at the quadrant intersection, where the four wavefronts meet. Figure \ref{riemann_hyb} (a) shows the density contours for the strong 2D Riemann problem with the LAD-C6 scheme, showing the problematic cells prior to the onset of the divergence. Further investigation revealed that the issue originated from implicit filtering applied to conservative variables during the fourth stage of each Runge–Kutta iteration. While this filtering is generally intended to stabilise the solution, its implicit nature can introduce non-physical oscillations in regions with sharp discontinuities, leading to a non-physical flow field. To mitigate this issue, filtering was disabled during the initial iterations to allow the flow to develop, and subsequently re-enabled (after five iterations) once the solution progressed beyond the initial transient phase. This approach (scheme referred to as LAD-C6$^*$) resulted in a stable and physically consistent flow field for the previously problematic strong Riemann configuration. 

\begin{algorithm} 
\caption{Main steps of the COMPSQUARE solver algorithm with LAD-GBR hybrid scheme}
\begin{algorithmic}[1]
\State Allocate variables, compute grid metric, initialisation and numerics 
\For{iter = 1 to $N_{\text{steps}}$} \Comment{Time iterations}
    \For{step = 1 to $N_{\text{rk}}$} \Comment{Runge-Kutta stages}
        \If{troubled\_cells $\neq$ 1}
        \State \textsc{LAD-based solver} ()
        \Else
        \State \textsc{GBR-based solver} ()
        \EndIf
    \EndFor
\EndFor
\State \textsc{Output} (Save required statistics to file)
\end{algorithmic}  
\label{hybrid_algo}
\end{algorithm}

\begin{figure}
\centering
\includegraphics[scale = 0.8]{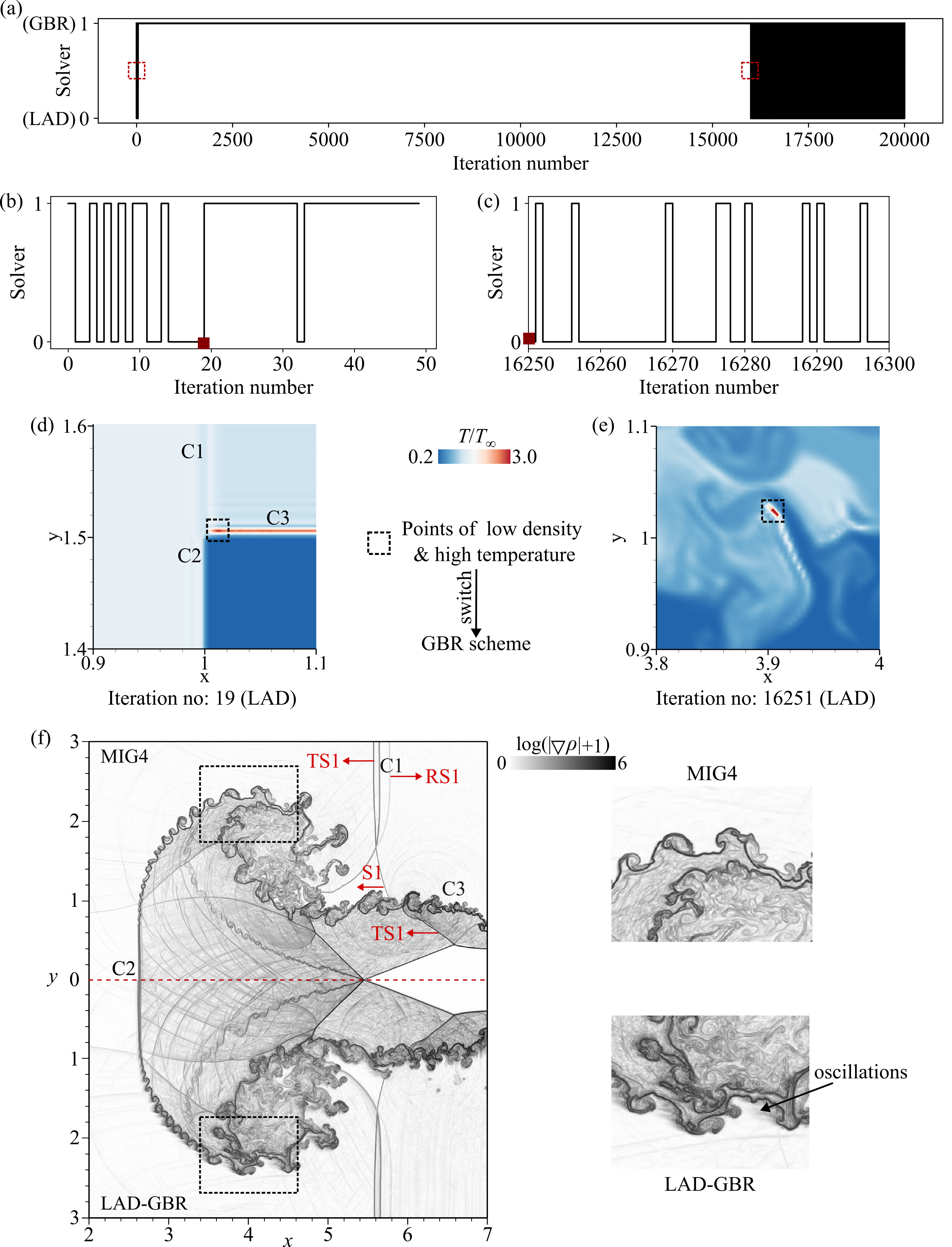}
    \caption{Compressible triple point problem - (a) switching history for the 
    entire simulation time, zoomed in region of the of the solver transition from (b) iterations 1 to 50 and (c) iterations 16250 to 16300; temperature contours with highlighted regions of troubled cells before switching to the GBR solver at iteration (d) 19 and (e) 16251; contours of density gradient magnitude shown at $t=5.0$ units, Top: MIG4 and bottom: hybrid LAD-GBR scheme.}
    \label{comp3p_switch_hybrid}
\end{figure}

\begin{figure}
    \centering
\includegraphics[scale=0.9]{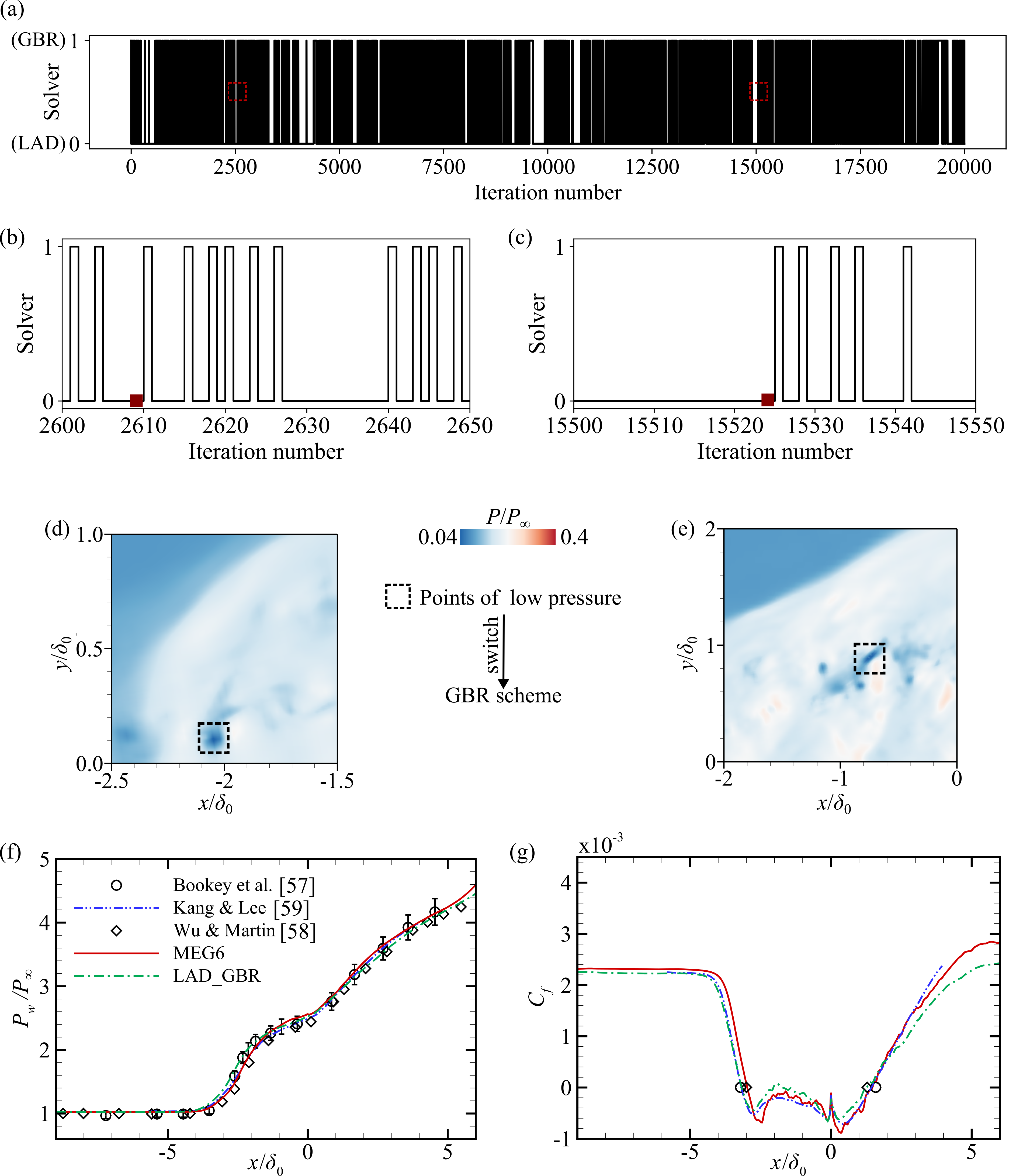}
    \caption{M2.9, $24^{\circ}$ ramp - (a) switching history for one flow-through time, (b) zoomed in region of the solver transition from iterations 2600 to 2650, (c) zoomed in region of the solver transition from iterations 15500 to 15550; temperature contours with highlighted regions of troubled cells before switching to the GBR solver at iteration (d) 2610 and (e) 15525; Profiles of streamwise variation of (f) wall pressure, and (g) skin friction coefficient computed using the hybrid LAD-GBR scheme and compared against the reference data.}
    \label{M2.9_switch_hybrid}
\end{figure}

Figure \ref{riemann_hyb} (b-d) illustrates the density contours of the strong Riemann problem with 30 uniformly spaced contour levels flooded by lines with the LAD-C6$^*$, MEG6 and MIG4 schemes. The scheme captures jet-like structures along with small-scale vortical features. However, the jet-like structure with LAD-C6$^*$ is less pronounced compared to the MIG4 and MEG6 schemes. Moreover, initial interfaces from initialisation are still present in the flow field obtained with the LAD-C6$^*$ scheme. The following strategy is further applied to the compressible triple point problem, but it failed to produce a stable solution. The divergence is once again observed at the interface of the contact discontinuity (C3). To this end, the use of explicit filtering instead of implicit filtering may enhance stability and prevent divergence in the flow field. Although the proposed formulation $\kappa^*$ in the LAD approach identifies regions of contact discontinuity, it was ineffective in mitigating the instabilities arising from strong contact discontinuities. Additional stabilisation techniques are necessary to properly resolve the contact discontinuity and ensure robust and reliable simulations.

While the LAD approach is computationally efficient, we have observed that it can lead to unstable solutions for test cases involving strong discontinuities. In contrast, gradient-based reconstruction (GBR) methods are robust and accurate but more expensive compared to LAD. A hybrid approach that leverages the strengths of both methods offers a promising solution. The proposed hybrid solver combines LAD and GBR schemes to address high-speed turbulent flows involving shock wave–boundary layer interactions. The findings presented in Section \ref{results} revealed that the test cases with the LAD-C6 scheme eventually diverged after encountering negative density and pressure values at a few grid points, leading to abrupt divergence. To overcome this, the proposed approach switches to the GBR-based scheme whenever near-vacuum pressure or a very small density is detected in the flow field. This improves stability without significantly increasing computational cost. In the actual hybrid implementation reported in the literature \cite{fernandez2018posteriori, bernardini2021streams, adler2019flow, soldati2024flew}, the scheme is switched only for troubled cells. However, for ease of implementation in the present study, the entire flow field is computed using the GBR scheme once troubled cells are identified. The pseudocode of the hybrid LAD-GBR framework is described in Algorithm \ref{hybrid_algo}, while the detailed steps of the individual LAD and GBR algorithms are highlighted in Algorithms \ref{LAD_algo} and \ref{GBR_algo}, respectively.

The hybrid solver employs explicit fourth-order central differencing (E4) for spatial discretisation with the LAD scheme, whereas the GBR solver uses the MEG6 scheme. The E4 scheme was selected instead of C6 for the LAD approach, as it provided greater stability within the current hybrid framework. A third-order explicit total variation diminishing (TVD) Runge-Kutta is employed for time marching. The efficacy of the hybrid solver is first demonstrated on the compressible triple point problem, which previously diverged with the LAD scheme. The setup of the test case is described in subsection \ref{comp3p_case}. Figure \ref{comp3p_switch_hybrid} (a) illustrates the switching history for the hybrid solver throughout the simulation. The LAD and GBR schemes are indicated as 0 and 1, respectively. During the simulation period, the LAD scheme is used for 13.6\% of the time, while the GBR scheme accounts for the remainder. Figures \ref{comp3p_switch_hybrid} (b) and (c) show the zoomed-in region of the solver transition for two different iteration ranges. Figures \ref{comp3p_switch_hybrid} (d) and (e) show the temperature contours with highlighted regions of troubled cells before switching to the GBR solver at iterations 19 and 16251. During initial iterations (Fig. \ref{comp3p_switch_hybrid} (d)), the flow field predicted using the LAD scheme exhibits a sharp temperature increase and a decrease in density along the contact discontinuity C3. This triggers repeated switches between the LAD and GBR schemes. The hybrid solver then stabilises with the GBR scheme, as the low-density region generated during the initial LAD phase persists and prevents a switch back to LAD. The flow solution is advanced with the GBR scheme until the shock wave S1 reflects into the domain (Fig. \ref{comp3p_switch_hybrid} (f)), at which the low-density region gradually dissipates and the hybrid framework dynamically adjusts between LAD and GBR depending on the local flow conditions. The solver switches to the GBR scheme when the local density, $\rho_l < 0.5 \rho_{C3}$, where $\rho_{C3}$ is the initial density in quadrant C3 (which is minimum among all quadrants), else, the hybrid solver continues with the LAD scheme. A density-based criterion is used, as divergence occurs at contact discontinuities where the pressure remains nearly constant, making the pressure threshold unsuitable for switching. The threshold is chosen based on observations of density variations in the MIG4 flow-field simulation, ensuring that the switching remains consistent and reliable. Figure \ref{comp3p_switch_hybrid} (f) illustrates the contours of density gradient magnitude and a magnified region of the same at $t=5$ units for the compressible triple point problem. The results obtained using MIG4 (top) and the hybrid LAD–GBR scheme (bottom) are shown. The results from the hybrid scheme exhibit marginal oscillations, whereas MIG4 captures the contact discontinuity without oscillations and with a sharper material interface.


The second test case is M2.9, $24^{\circ}$ compression ramp test case, which previously diverged within the LAD framework. Figure \ref{M2.9_switch_hybrid} (a) illustrates the switching history for the hybrid solver for one flow-through time, with the LAD and GBR schemes indicated as 0 and 1, respectively. The hybrid solver switches from LAD to the GBR scheme when the local pressure drops below half the freestream value and switches back to LAD otherwise. During the total simulation time, the LAD solver is active for 82.4\% of the duration. Hence, the hybrid solver achieves a substantial computational speedup of $1.67\times$ compared to the standalone C-GBR solver with the MEG6 scheme. Figures \ref{M2.9_switch_hybrid} (b) and (c) show the zoomed-in region of the solver transition for two different iteration ranges. Figures \ref{M2.9_switch_hybrid} (d) and (e) show the pressure contours at two different iterations marked in (b,c). The inset boxes highlight the regions of troubled cells using LAD prior to switching to the GBR solver. Figure \ref{M2.9_switch_hybrid} (f) shows the streamwise variation of the wall pressure. The present LES simulations using the hybrid solver (LAD-GBR) and MEG6 are compared with the DNS results \cite{ wu2007direct, kokkinakis2020direct} and the experimental data of Bookey et al. \cite{bookey2005new}. The results obtained from both methods are in good agreement with previously reported studies, although marginal differences are evident in the LAD–GBR approach near the onset of the separation bubble. Figure \ref{M2.9_switch_hybrid} (g) plots the streamwise evolution of the skin-friction coefficient. Separation and reattachment points from the data of Bookey et al. \cite{bookey2005new} and Wu \& Martin \cite{wu2007direct} are also indicated. The results obtained using the MEG6 scheme show good agreement with both the experimental and DNS data, while the LAD-GBR approach exhibits a noticeable underprediction in $C_f$ after reattachment. 

It is worth noting that the hybrid solver implementation demonstrated in this work is not optimised for efficiency. The hybrid solver switches from the LAD to the GBR scheme for the entire flow field. Instead, the hybrid solver can be further improved by adopting the numerical framework similar to that of Fidalgo et al. \cite{fernandez2018posteriori}, where the scheme can be switched from LAD to GBR only across the troubled cells.




\section{Conclusions} \label{summary}

We have demonstrated the performance of two promising methods for capturing high-speed flows, local artificial diffusivity (LAD-C6) and centralised gradient-based reconstruction (C-GBR: MEG6/MIG4). The former method is computationally economical, while the latter is more robust, but expensive. The LAD approach is based on adding grid-dependent artificial fluid transport coefficients to capture sharp discontinuities. In contrast, the C-GRB scheme exploits the characteristic variable transformation to selectively treat each characteristic wave, yielding cleaner results, and employs an HLLC approximate Riemann solver to compute the inviscid fluxes. The efficacy of these schemes is first demonstrated on 1D and 2D single-species benchmark cases. C-GBR shows superior results, while LAD suffers from numerical dissipation and marginal oscillations. Further evaluations are carried out to examine shock-capturing capabilities in 3D compressible turbulent flows. Although both schemes deliver competitive results for LES of supersonic turbulent boundary layers, the LAD schemes fail to maintain stability in supersonic and hypersonic compression ramp simulations. In contrast, C-GBR successfully handles all test cases, albeit at a higher computational cost.

We further compare the computational efficiency of both approaches across a single-core Intel Xeon Gold 6240R and an NVIDIA A100 GPU with 80 GB HBM2e memory. A directive-based OPENACC-based GPU parallelism is exploited. On GPUs, LAD-C6 demonstrates a speed-up of $1.17-1.22\times$ and $1.38-1.43\times$ compared to the MEG6 and MIG4 schemes, while LAD-E4 is $1.9-2.0\times$ faster than MEG6 and $2.2-2.3\times$ faster than MIG4. However, both the MEG6 and MIG4 schemes show higher GPU acceleration compared to the LAD schemes, achieving $2\times$ speed-ups relative to their CPU runtimes, thereby demonstrating their efficiency in GPU architectures. The subroutine-level breakdown
on the GPU reveals that viscous routines contribute to 57.5\% of the total runtime for the LAD-C6 scheme, while the inviscid routine accounts for the largest share, 70.4\%, in the MIG4 implementation.

A hybrid LAD-GBR framework is proposed that leverages the strengths of both LAD and GBR schemes. The efficacy of the implementation is demonstrated on a compressible triple point problem and a supersonic compression ramp. For the M2.9 $24^{\circ}$ test case, the hybrid solver achieves a substantial computational speed-up of $1.67\times$ relative to the standalone C-GBR with the MEG6 scheme.

\section*{Acknowledgements}
The authors thank the National PARAM Supercomputing Facility (NPSF) for providing computing resources in the PARAM Siddhi cluster under the National Supercomputing Mission. We also acknowledge NSM providing access to the computing resources of ‘PARAM RUDRA’ at P G Senapathy Center, Play Field Ave, Indian Institute of Technology, Chennai, Tamil Nadu 600036, which is implemented by C-DAC and supported by the Ministry of Electronics and Information Technology (MeitY), and Department of Science and Technology (DST), Government of India.



\printcredits

\bibliographystyle{unsrt}

\bibliography{cas-refs}


\end{document}